\documentclass
[superscriptaddress,secnumarabic,amssymb,amsmath,nobibnotes,aps,prd,showkeys,showpacs,nofootinbib]{revtex4}%
\usepackage{graphicx}
\usepackage{epsf}
\usepackage{bm}
\usepackage{amsmath}
\usepackage{amsfonts}
\usepackage{amssymb}
\usepackage{epstopdf}%
\providecommand{\U}[1]{\protect\rule{.1in}{.1in}}
\newcommand{\be}{\begin{equation}}
\newcommand{\ee}{\end{equation}}

\newcommand{\mincir}{\raise
-3.truept\hbox{\rlap{\hbox{$\sim$}}\raise4.truept\hbox{$<$}\ }}
\newcommand{\magcir}{\raise
-3.truept\hbox{\rlap{\hbox{$\sim$}}\raise4.truept\hbox{$>$}\ }}

\begin{document}
\title{Dynamical analysis in scalar field cosmology}
\author{Andronikos Paliathanasis}
\email{paliathanasis@na.infn.it}
\affiliation{Dipartimento di Fisica, Universita' di Napoli, ``Federico II'', Complesso
Universitario di Monte S. Angelo, Via Cinthia 9, I-80126, Napoli, Italy}
\affiliation{Istituto Nazionale di Fisica Nucleare (INFN) Sez. di Napoli, Complesso
Universitario di Monte S. Angelo, Via Cinthia 9, I-80126 Naples, Italy}
\author{Michael Tsamparlis}
\email{mtsampa@phys.uoa.gr}
\affiliation{Faculty of Physics, Department of Astrophysics - Astronomy - Mechanics,
University of Athens, Panepistemiopolis, Athens 157 83, Greece}
\author{Spyros Basilakos}
\email{svasil@academyofathens.gr}
\affiliation{Academy of Athens, Research Center for Astronomy and Applied Mathematics,
Soranou Efesiou 4, 11527, Athens, Greece}
\author{John D. Barrow}
\email{jdb34@hermes.cam.ac.uk}
\affiliation{DAMTP, Centre for Mathematical Sciences, University of Cambridge, Wilberforce
Rd., Cambridge CB3 0WA, UK}

\begin{abstract}
We give a general method to find exact cosmological solutions for scalar-field
dark energy in the presence of perfect fluids. We use the existence of
invariant transformations for the Wheeler De Witt (WdW) equation. We show that
the existence of a point transformation under which the WdW equation is
invariant is equivalent to the existence of conservation laws for the field
equations, which indicates the existence of analytical solutions. We extend
previous work by providing exact solutions for the Hubble parameter and the
effective dark-energy equation of state parameter for cosmologies containing a
combination of perfect fluid and a scalar field whose self-interaction
potential is a power of hyperbolic functions. We find solutions explicity when
the perfect fluid is radiation or cold dark matter and determine the effects
of non-zero spatial curvature. Using the Planck 2015 data, we determine the
evolution of the effective equation of state of the dark energy. Finally, we
study the global dynamics using dimensionless variables. We find that if the
current cosmological model is Liouville integrable (admits conservation laws)
then there is a unique stable point which describes the de-Sitter phase of the universe.

\end{abstract}

\maketitle

\section{Introduction}

The discovery of the accelerated expansion of the universe (see \cite{Teg04}
and references therein) has opened a new window in cosmological studies.
Indeed, the underlying physical process responsible for this phenomenon is
considered as one of the fundamental problems in cosmology. Within the
framework of general relativity, scalar fields provide possible dark energy
models which can describe, but not so far explain, this acceleration. Scalar
field models require the choice of a self-interaction potential $V(\phi)$ for
the scalar field $\phi$. The considerations of a specific potential $V(\phi)$
is done by an adhoc requirement (ansatz). In this manner various candidates
have been proposed in the literature, such as an inverse power law,
exponential, hyperbolic and the list goes on (for review see \cite{Ame10} and
references therein). One such ansatz, which we shall consider in the
following, has been done by Rubano and Barrow \cite{Barrow} (see also
\cite{Urena}) who found that if the scalar field behaves as a perfect fluid
then the potential $V(\phi)\propto\mathrm{sinh}^{p}(q\phi)$, where the
constants $q,p$ are given in terms of observable cosmological parameters,
namely the dark-energy equation of state (EoS) parameter and $\Omega_{m0}$.

Obviously it is important to proposed potentials which are realistic and at
the same time lead to exactly soluble models in the FLRW spacetime. The reason
for employing the Noether symmetries is that Noether symmetries provide us
with the Noether integrals which facilitate the analytic solution of the field
equations. As an example, in extended theories of gravity, where the Birkhoff
theorem is not guaranteed, the Noether approach provides a means of describing
the global dynamics using the first integrals of motion \cite{CapFru}.
Furthermore, besides the technical possibility of reducing the dynamical
system, the first integrals of motion always give rise to conserved currents
that are not only present in physical space-time but also in configuration
spaces (see discussion in \cite{Capp96}). In space-time such currents are
linear (momentum, angular momentum etc.) but in configuration space the
conserved quantities appear as relations between dynamical variables
\cite{Ritis}. The latter implies that in the configuration space
(minisuperspace), the first integrals are considered as "selection rules" for
scalar field potentials and coupling functions in the case of modified gravity
theories. The above features have inspired many authors to propose the
admittance of a Noether (point or dynamical) symmetry by the field equations
as a selection rule for dark-energy models, including those of modified
gravity
\cite{Cap96,Dong,Darabi,Wei,Kucu,CapHam,Vak,Dim,Dim2,Paliathanasis,BasFT,DynSym}%
.\textbf{ }
%On the other hand it has been shown that Noether
%symmetries are related to the geometry of the space where the dynamical system
%"move" \cite{Paliathanasis}.

In particular it has been shown that the dynamical Noether symmetries are
associated with the Killing tensors of the space \cite{DynSym,Kalotas} and
that the Noether point symmetries are related to the homothetic algebra of the
space (see \cite{TsamGRG} and references therein). Therefore the requirement
of a Noether symmetry is indeed a geometric demand hence independent of the
particular dynamics.

In scalar field cosmology the system develops in minisuperspace whose geometry
is defined by the field equations. Therefore the requirement that the field
equations admit a Noether symmetry becomes, according to the above results, a
geometric requirement on a geometry which is inherent to the system.

In this work by using a more general geometric criterion, i.e. by employing
the Lie point symmetries of the Wheeler-DeWitt (WdW) equation, we extend the
work of Rubano and Barrow \cite{Barrow} to a general family of hyperbolic
scalar-field potentials $V(\phi)$.

In section \ref{SFC}, we give the basic theory of the scalar-field cosmology
in a FLRW spacetime. The basic definitions and results from the Lie and
Noether point symmetries of partial differential equations and the application
in the WdW equation are presented in section \ref{preliminaries}. In section
\ref{WDWsym}, we consider our cosmological model, which includes a hyperbolic
family of scalar field potentials with a perfect fluid with constant equation
of state parameter $w_{m}$. We study the existence of Lie point symmetries of
the WdW equation, where we find that for our model the WdW equation admits Lie
point symmetries if the free parameters of the potential and the parameter
$w_{m}$ are related. In section \ref{ExactSol} we apply the Lie point
symmetries of the WdW equation in order to construct invariant solutions of
the WdW equation and exact solutions of the field equations. Also, in section
\ref{dynanalysis} we perform a dynamical analysis by studying the fixed points
of the field equations in the dimensionless variables for the general model
and we show that when the cosmological model is Liouville integrable (the
model admits conservation laws), there is a unique stable point which
describes the de Sitter universe. Finally, in section \ref{conclusion} we draw
some conclusions.

\section{Scalar-field cosmology}

\label{SFC} We start with the Friedmann-Lema\^{\i}tre-Robertson-Walker (FLRW)
spacetime with line element ($c\equiv1$)
\begin{equation}
ds^{2}=-dt^{2}+a^{2}(t)\frac{1}{(1+\frac{\mathit{K}}{4}\mathbf{x}^{2})^{2}%
}(dx^{2}+dy^{2}+dz^{2}). \label{SF.1}%
\end{equation}
The total action of the field equations is written as
\begin{equation}
S=S_{EH}+S_{\phi}+S_{m}, \label{SF.01}%
\end{equation}
where $S_{EH}=\int dx^{4}\sqrt{-g}R~$is the Einstein-Hilbert action, $R$ is
the Ricci scalar of the underlying space, $S_{\phi}$ is the action of the
scalar field
\begin{equation}
S_{\phi}=\int dx^{4}\sqrt{-g}\left[  -\frac{1}{2}g^{\mu\nu}\phi_{;\mu}%
\phi_{;\nu}+V(\phi)\right]  , \label{SF.02}%
\end{equation}
and $S_{m}=\int dx^{4}\sqrt{-g}L_{m}$ is the matter term. We assume that
$\phi$ inherits the symmetries of the metric (\ref{SF.1}) therefore $\phi(t)$
and consequently $\phi_{;\nu}=\dot{\phi}\delta_{\nu}^{0}$ where $\dot{\phi
}=\frac{d\phi}{dt}$.

From the action (\ref{SF.01}), we have the Einstein field equations
\cite{Ame10}%
\begin{equation}
R_{\mu\nu}-\frac{1}{2}g_{\mu\nu}R=\kappa\tilde{T}_{\mu\nu} \label{SF.03}%
\end{equation}
where $\kappa=8\pi G\equiv1$, $R_{\mu\nu}$ is the Ricci tensor and $\tilde
{T}_{\mu\nu}$ is the total energy momentum tensor given by $\tilde{T}_{\mu\nu
}\equiv T_{\mu\nu}+T_{\mu\nu}(\phi)$. \ $T_{\mu\nu}$ is the energy-momentum
tensor of baryonic\ matter and radiation and $T_{\mu\nu}(\phi)$ is the
energy-momentum tensor associated with the scalar field $\phi$. Modeling the
expanding universe as a fluid (which includes radiation, matter and DE) with
$4-$velocity $u_{\mu}$, proper isotropic density $\rho_{m}$ and proper
isotropic pressure $P_{m}$ gives $\tilde{T}_{\mu\nu}=-P\,g_{\mu\nu}%
+(\rho+P)u_{\mu}u_{\nu}$, where $\rho=\rho_{m}+\rho_{\phi}$ and $P=P_{m}%
+P_{\phi}$. The variable $\rho_{\phi}$ denotes the energy density of the
scalar field and $P_{\phi}$ is the corresponding isotropic pressure. Moreover
the parameters $(\rho_{\phi},P_{\phi})$ of the scalar field are given by
\begin{equation}
\rho_{\phi}\equiv\frac{1}{2}\dot{\phi}^{2}+V(\phi)~,~P_{\phi}\equiv\frac{1}%
{2}\dot{\phi}^{2}-V(\phi). \label{SF.04}%
\end{equation}

For the FLRW spacetime, for comoving observers ($u^{\mu}=\delta_{0}^{\mu}$),
the Einstein field equations (\ref{SF.03}) are
\begin{equation}
H^{2}=\frac{\kappa}{3}\left(  \rho_{m}+\rho_{\phi}\right)  -\frac{\mathit{K}%
}{a^{2}} \label{SF.07}%
\end{equation}
and
\begin{equation}
3H^{2}+2\dot{H}=-\kappa(P_{m}+P_{\phi})-\frac{\mathit{K}}{a^{2}},
\label{frie2}%
\end{equation}
where $H(t)\equiv\dot{a}/a$ is the Hubble function.

Furthermore, assuming that the scalar field and matter do not interact, we
have the\ two conservation laws%
\begin{equation}
\dot{\rho}_{m}+3H(\rho_{m}+P_{m})=0\, \label{SF.08}%
\end{equation}%
\begin{equation}
\dot{\rho}_{\phi}+3H(\rho_{\phi}+P_{\phi})=0\, \label{SF.09}%
\end{equation}
while the corresponding equation of state (EoS) parameters are given by
$w_{m}=P_{m}/\rho_{m}$ and $w_{\phi}=P_{\phi}/\rho_{\phi}$. In what follows we
assume a constant $w_{m},$ so that $\rho_{m}=\rho_{m0}a^{-3(1+w_{m})}$
($w_{m}=0$ for cold matter and $w_{m}=1/3$ for \ radiation), where $\rho_{m0}$
is the matter density at the present time. Generically, some high-energy field
theories suggest that the dark energy EoS parameter may be a function of
cosmic time (see, for instance, \cite{Ellis05}).

Replacing (\ref{SF.04}) in (\ref{SF.09}) we have the Klein-Gordon equation{{%
\begin{equation}
\ddot{\phi}+3H\dot{\phi}+V_{,\phi}=0 \label{SF.10}%
\end{equation}
where }}$V_{,\phi}=\frac{dV}{d\phi}$. Furthermore the corresponding dark
energy EoS parameter is
\begin{equation}
w_{\phi}=\frac{P_{\phi}}{\rho_{\phi}}=\frac{(\dot{\phi}^{2}/2)-V(\phi)}%
{(\dot{\phi}^{2}/2)+V(\phi)}%
\end{equation}
which means that when $w_{\phi}<-\frac{1}{3}$ then $\dot{\phi}^{2}<V(\phi)$.
On the other hand, if the kinetic term of the scalar field is negligible with
respect to the potential energy, i.e. $\frac{\dot{\phi}^{2}}{2}\ll V(\phi)$,
then the equation of state parameter is $w_{\phi}\simeq-1$.

From the above analysis it becomes clear that the unknown quantities of the
problem are $a(t),$ $\phi(t)$ and $V(\phi)$ whereas we have only two
independent differential equations available namely Eqs. (\ref{frie2}) and
(\ref{SF.10}). Therefore, in order to solve the system of differential
equations we need to make an additional assumption (ansatz). This usually
concerns the functional form of the scalar field potential $V(\phi)$. In the
literature, due to the unknown nature of DE, there are many forms of this
potential which describe differently the physical features of the scalar field
(for instance see
\ \cite{Ame10,Barrow,Urena,Ratra88,Sievers,Bertacca,Brax,Gorini,Frieman,Sahni,BarrowLog}%
).

As far as the exact solution of the field equations (\ref{SF.07}),
(\ref{frie2}) and (\ref{SF.10}) is concerned there are few solutions with
spatial curvature \cite{Halliwel,Easther} and even fewer solutions are known
for a perfect fluid and a scalar field
\ \cite{DynSym,BarrowS,Chimento,Chimento2,Ester2}. A special solution for a
spatially flat FLRW spacetime ($\mathit{K}=0$) which contains a perfect
fluid\ with a constant equation of state parameter $P_{m}=\left(
\gamma-1\right)  \rho_{m}$ and a scalar field with a constant equation of
state parameter $w_{\phi}=\gamma_{\phi}-1=P_{\phi}/\rho_{\phi}$,
%$P_{\phi}=w_{\phi}(a)\rho_{\phi}=\left(  \gamma_{\phi}-1\right)  \rho_{\phi}~$
has been found in \cite{Barrow}. Specifically in \cite{Barrow} it has been
shown that under these assumptions one solves the field equations and finds
the potential $V\left(  \phi\right)  $:
\begin{equation}
V\left(  \phi\right)  =3H_{0}^{2}\left(  1-\Omega_{m0}\right)  \left(
1-\frac{\gamma_{\phi}}{2}\right)  \left(  \frac{1-\Omega_{m0}}{\Omega_{m0}%
}\right)  ^{\frac{\gamma_{\phi}}{\gamma-\gamma_{\phi}}}\left[  \sinh\left(
\sqrt{3}\frac{\gamma-\gamma_{\phi}}{\sqrt{\gamma_{\phi}}}\left(  \phi-\phi
_{0}\right)  \right)  \right]  ^{-\frac{2\gamma_{\phi}}{\gamma-\gamma_{\phi}}%
}. \label{Bgp1}%
\end{equation}
Evidently, this solution is a special solution, in the sense that it exists
for specific initial conditions, for example with $w_{\phi}\left(  z\right)  $ constant.

In the following we consider a spatially-flat FLRW spacetime with a perfect
fluid $P_{m}=\left(  \gamma-1\right)  \rho_{m}$ and a scalar field and assume
that the potential has the generic form
%\begin{subequations}%
\begin{equation}
V\left(  \phi\right)  =V_{0}\left[  \alpha\cosh\left(  p\phi\right)
+\beta\sinh\left(  p\phi\right)  \right]  ^{q}, \label{Bgp}%
\end{equation}
where $V_{0},\alpha,\beta,p,$and $q$ are constants. This potential (\ref{Bgp})
is a generalization of (\ref{Bgp1}) and our aim is to determine exact
solutions of the field equations for a particular relation between the
constants $p,q$ of the potential (\ref{Bgp}) and the barotropic parameter
$\gamma$ of the perfect fluid.

\section{Preliminaries}

\label{preliminaries}

In this section, we show that if the WdW equation admits Lie symmetries which
form an Abelian Lie algebra, then the WdW equation is Liouville integrable;
that is, the field equations can be solved by quadratures.

\subsection{Lie and Noether point symmetries}

A Lie symmetry of a differential equation $H=H(x^{i},u^{A},u_{,i}^{A}%
,u_{,ij}^{A})$ is the generator of the one parameter point transformation
which leaves invariant the differential equation $H$. That means that if
$\mathbf{X=}\xi^{i}(x^{k},u^{B})\partial_{i}+\eta^{A}(x^{k},u^{B}%
)\partial_{\eta^{A}}$ is a Lie symmetry for $H,$ then there exists a function
$\lambda$ such that the following condition holds \cite{Bluman,Ibrag}
\begin{equation}
\mathbf{X}^{[2]}(H)=\lambda H~,~\mathbf{mod}H=0
\end{equation}
where $\mathbf{X}^{[2]}=\mathbf{X}+\eta_{i}^{A}\partial_{u_{,i}^{A}}+\eta
_{ij}^{A}\partial_{u_{,ij}^{A}}\mathbf{,~}$is the second prolongation
vector\footnote{Where $\eta_{ij...,j_{n}}^{A}=D_{j_{n}}\left(  \eta
_{ij...j_{n-1}}^{A}\right)  -u_{,ij...j_{n-1}k}^{A}D_{j}\left(  \xi
^{k}\right)  $ and $D_{i}$ is the total derivative.} of $\mathbf{X}$.

The importance of Lie symmetries is that each symmetry can be used to reduce
the number of dependent variables.\ Solutions which follow from the
application of Lie symmetries are called invariant solutions.

For differential equations which arise from a variational principle there
exists a special class of Lie symmetries, the Noether symmetries. Noether
symmetries are Lie symmetries which leave the action integral invariant.
According to Noether's theorem to each Noether symmetry there corresponds a
conserved Noether current.

The condition for a Noether symmetry is that there exists a vector field
$A^{i}=A^{i}(x^{i},u)$ such that the following condition is satisfied:
\begin{equation}
\mathbf{X}^{[1]}L+LD_{i}\xi^{i}=D_{i}A^{i}~. \label{pr.08}%
\end{equation}
The corresponding Noether current $I^{i}$ is defined by the expression
\begin{equation}
I^{i}=\xi^{k}\left(  u_{k}^{A}\frac{\partial L}{\partial u_{i}^{A}}-L\right)
-\eta^{A}\frac{\partial L}{\partial u_{i}^{A}}+A^{i}~. \label{pr.09}%
\end{equation}
and it is conserved, that is, it satisfies the relation~\ $D_{i}I^{i}=0$
\cite{Bluman}.

As we discussed above, the method of using the Noether symmetries of the
cosmological field equations has been applied by many authors in scalar-field
cosmology, in $f(R)$ gravity and other modified gravity theories. Recently
\cite{AnIJGMMP,anthesis}, it has been proposed that the cosmological model
will be determined by the existence of Lie symmetries of the WdW equation of
quantum cosmology.

This selection rule is more general than that imposed by the Noether
symmetries of the field equations, because, as it has been shown in
\cite{anthesis}, the WdW equation is possible to admit Lie point symmetries
while the Lagrangian of the field equations does not admit Noether point
symmetries. In the following we discuss the application of Lie symmetries in
the WdW equation. Specifically, we discuss the reduction process and we show
how to construct Noetherian conservation laws for a conformally related
Lagrangian of the field equations from the Lie point symmetries of the WdW equation.

\subsection{Minisuperspace and invariant solutions of the WdW equation}

The Lagrangian of the field equations in minimally coupled scalar field
cosmology in \ a spatially flat FLRW spacetime with a perfect fluid with a
constant equation of state parameter $P_{m}=\left(  \gamma-1\right)  \rho_{m}$
is
\begin{equation}
L\left(  a,\dot{a},\phi,\dot{\phi}\right)  =-3a\dot{a}^{2}+\frac{1}{2}%
a^{3}\dot{\phi}^{2}-a^{3}V\left(  \phi\right)  -\rho_{m0}a^{-3\left(
\gamma-1\right)  }. \label{lang.01}%
\end{equation}
The field equations are the Euler-Lagrange equations of (\ref{lang.01}) with
respect to the variables $\left(  a,\phi\right)  $ and are equations
(\ref{frie2}) and (\ref{SF.10}). As the Lagrangian is independent of time, we
also have the Hamiltonian constraint (\ref{SF.07}), which, in terms of the
momenta $p_{a}=\frac{\partial L}{\partial\dot{a}},~p_{\phi}=\frac{\partial
L}{\partial\dot{\phi}},$ becomes
\begin{equation}
-\frac{1}{12a}p_{a}^{2}+\frac{1}{2a^{3}}p_{\phi}^{2}+a^{3}V\left(
\phi\right)  +\rho_{m0}a^{-3\left(  \gamma-1\right)  }=0. \label{lang.02}%
\end{equation}
Finally, the field equations are equivalent to the following Hamiltonian
system:%
\[
\dot{a}=-\frac{1}{6a}\dot{p}_{a}~,~\dot{\phi}=\frac{1}{a^{3}}p_{\phi}%
~,~\dot{p}_{\phi}=-a^{3}V_{,\phi},
\]%
\[
\dot{p}_{a}=-\frac{1}{12}\frac{p_{a}^{2}}{a^{2}}+\frac{3}{2}\frac{p_{\phi}%
^{2}}{a^{4}}-3a^{2}V\left(  \phi\right)  +\left(  3\gamma-3\right)  \rho
_{m0}a^{-3\gamma+2}=0.
\]

The WdW equation is the Klein Gordon equation which is defined by the
conformal Laplacian operator. The general conformal Klein Gordon equation is:%
\begin{equation}
\Delta\Psi+\frac{n-2}{4\left(  n-1\right)  }R\left(  x^{k}\right)
\Psi+V_{eff}\left(  x^{k}\right)  \Psi=0,\label{lang.03}%
\end{equation}
where $\Delta=\frac{1}{\sqrt{\left\vert g\right\vert }}\frac{\partial
}{\partial x^{i}}\left(  \sqrt{\left\vert g\right\vert }\frac{\partial
}{\partial x^{j}}\right)  $ is the Laplacian operator, $g_{ij}$ is the metric
of the space and $n=\dim g_{ij}$. We brake the Lagrangian (\ref{lang.01}) in
two parts. The kinematic part which we consider as a Riemannian space with
dimension $n=2,$ and line element%
\begin{equation}
d{\hat{s}}^{2}=-6ada^{2}+a^{3}d\phi^{2}\label{lang.04}%
\end{equation}
which we call the minisuperspace, and the dynamic part which is defined by the
potential $V_{eff}\left(  a,\phi\right)  =2a^{3}\left[  V\left(  \phi\right)
+\rho_{m0}a^{-3\gamma}\right]  $. Specifically, the minisuperspace is a
2-dimensional Lorentz manifold whose coordinates are the scale factor and the
scalar field. The symmetries of this space are related to the Lie and the
Noether symmetries of the dynamical field equations. The metric of the
minisuperspace is defined by the kinematic part of the Lagrangian
(\ref{lang.01}) for the dynamics of the field. The rest of the Lagrangian is
considered to be the "effective potential" of the dynamical system that is
defined by the gravitational and scalar fields. Therefore, using the
minisuperspace (\ref{lang.04}) the WdW equations becomes
\begin{equation}
\Delta\Psi+2a^{3}\left[  V\left(  \phi\right)  +\rho_{m0}a^{-3\gamma}\right]
\Psi=0,\label{lang.05}%
\end{equation}
where the Laplacian operator $\Delta$ is given by%
\begin{equation}
\Delta\equiv-\frac{1}{6a}\left(  \frac{\partial^{2}}{\partial a^{2}}%
+\frac{\partial}{\partial a}\right)  +\frac{1}{a^{3}}\frac{\partial^{2}%
}{\partial\phi^{2}}.\label{lang.06}%
\end{equation}

In \cite{AnIJGMMP}, it was proved that the Lie point symmetries of equation
(\ref{lang.03}) are related to the conformal algebra of the minisuperspace
$g_{ij}$. More specifically, it has been shown that:

A. The general form of the Lie point symmetry vector is
\begin{equation}
\mathbf{X}=\xi^{i}\left(  x^{k}\right)  \partial_{i}+\left[  \frac{(2-n)}%
{2}\psi\Psi+a_{0}\Psi\right]  \partial_{\Psi}\mathbf{,} \label{lang.07}%
\end{equation}
where$~\xi^{i}\left(  x^{k}\right)  $ is a conformal Killing vector of the
minisuperspace, with conformal factor $\psi\left(  x^{k}\right)  .$

B. The Lie point symmetry condition which constrains the potential is
$\mathcal{L}_{\xi}V_{eff}+2\psi V_{eff}=0$.

We require now that equation (\ref{lang.03}) admits as Lie point symmetry the
vector (\ref{lang.07}). Then under the coordinate transformation
$x^{i}\rightarrow y^{i}$ so that $\xi^{i}\left(  x^{k}\right)  \partial
_{i}\rightarrow\partial_{J}$ , the Lie symmetry vector (\ref{lang.07}) becomes%
\begin{equation}
\mathbf{X}=\partial_{J}+\left[  \frac{2-n}{2}\psi\Psi+a_{0}\Psi\right]
\partial_{\Psi}. \label{lang.08}%
\end{equation}
There exist two equivalent methods to reduce the WdW equation by means of the
symmetry vector (\ref{lang.08}).

a) In the first method we calculate the zero-order invariants from the
Lagrange system ($b\neq J),$
\begin{equation}
\frac{dy^{b}}{0}=\frac{dy^{J}}{1}=\frac{d\Psi}{\left(  \frac{2-n}{2}\psi
+a_{0}\right)  \Psi}, \label{lang.09}%
\end{equation}
which turn out to be%
\begin{equation}
y^{b},~\Psi\left(  y^{b},y^{J}\right)  =\Phi\left(  y^{b}\right)  \exp\left[
\int\left(  \frac{2-n}{2}\psi+a_{0}\right)  dy^{J}\right]  . \label{lang.10}%
\end{equation}

b) The second method is to write the Lie point symmetry as a Lie B\"{a}cklund
symmetry. The Lie point symmetry (\ref{lang.08}) is equivalent to the contact
symmetry%
\begin{equation}
\bar{X}=\left(  \Psi_{J}-\left(  \frac{2-n}{2}\psi+a_{0}\right)  \Psi\right)
\partial_{\Psi}, \label{lang.11}%
\end{equation}
from which we obtain the differential equation%
\begin{equation}
\Psi_{J}-\left(  \frac{2-n}{2}\psi+a_{0}\right)  \Psi=a_{1}\Psi.
\label{lang.12}%
\end{equation}
We set $a_{0}+a_{1}=Q_{0}$ and find that the solution of the reduced equation
is%
\begin{equation}
\Psi\left(  y^{b},y^{J}\right)  =\Phi\left(  y^{b}\right)  \exp\left[
\int\left(  \frac{2-n}{2}\psi+Q_{0}\right)  dy^{J}\right]  \label{lang.13}%
\end{equation}
from which it follows again that the coordinate $y^{J}$ is factored out\ from
the solution of the wavefunction $\Psi\left(  y^{b},y^{J}\right)  .$

In the WKB approximation, $\Psi\left(  x^{k}\right)  \sim e^{iS\left(
x^{k}\right)  }$ the WdW equation reduces to a (null) Hamilton-Jacobi
equation. The latter can be seen as the Hamilton-Jacobi equation of a
Hamiltonian system moving in the same geometry under the conformal Laplace
operator of the WdW equation and with the same potential. Specifically, the
WdW equation (\ref{lang.05}) in scalar field cosmology provides the null
Hamilton-Jacobi equation:
\begin{equation}
-\frac{1}{12a}\left(  \frac{\partial S}{\partial a}\right)  ^{2}+\frac
{1}{2a^{3}}\left(  \frac{\partial S}{\partial\phi}\right)  ^{2}+a^{3}V\left(
\phi\right)  +\rho_{m0}a^{-3\left(  \gamma-1\right)  }=0. \label{lang.14}%
\end{equation}

Furthermore, in \cite{AnIJGMMP} it was also shown that the symmetries of the
WdW equation can be used in order to find Noether point symmetries for
classical particles. However, the null Hamilton-Jacobi equation is separable
if the $n$- dimensional Hamiltonian system admits $n$ conservation laws
$\Phi_{I}$ (symmetries) i.e. $n$ corresponding Noether symmetries which are
independent and in involution, i.e. $\left\{  \Phi_{I},\Phi_{K}\right\}  =0$
where $\left\{  .,.\right\}  $ denotes the Poisson bracket. If this is the
case, then the Hamiltonian system is Liouville integrable \cite{Arnold}. That
means that it is possible for the WdW equation to admit an invariant solution
and at the same time the classical Hamiltonian system to be not integrable.
Therefore, in order for the WdW equation to admit an invariant solution and
the Hamiltonian system to be Liouville integrable, the $n$- dimensional
WdW\emph{ }equation must admit at least $n-1$ independent Lie point
symmetries, $X_{I}$, which form an Abelian Lie algebra. If this is the case,
the zero-order invariants of these $n-1$ Lie point symmetries will give the
solution of the WdW equation in the form%
\begin{equation}
\Psi\left(  \bar{x}^{n},\bar{x}^{J}\right)  =\Phi\left(  \bar{x}^{n}\right)
\exp\left[  \sum_{J=1}^{n-1}\int\left(  \frac{2-n}{2}\psi-Q_{J}\right)
d\bar{x}^{J}\right]  \mathbf{,}%
\end{equation}
where~$Q_{J}$ are constants,~$J=1,2,...,n-1,$ and the function $\Phi\left(
\bar{x}^{n}\right)  $ satisfies a linear second-order ODE. That is, when the
field equations are Liouville integrable by Noether point symmetries then
there exists a coordinate system where the WdW equation admits $n$ oscillatory
terms in the solution and \textit{vice versa}. It is important to note that
this result is more general and includes the one given in \cite{CapHam} when
$\psi\left(  x^{k}\right)  =0;$ that is, if one considers the Killing algebra
of the minisuperspace only. We conclude that for the reduction/solution of the
WdW we may consider directly the Lie point symmetries of the WdW equation
which are given in terms of the CKVs of the space instead of restricting
ourselves to the Noether point symmetries only, as has been done in
\cite{CapHam}.

Below, we study the Lie point symmetries and the WdW equation for the
potentials of the form (\ref{Bgp}) which generalize the work done in
\cite{Barrow,Urena}.

\section{Lie point symmetries of the Wheeler-DeWitt equation}

\label{WDWsym}

We are considering a scalar field cosmological model which contains a
quintessence scalar field
%($\varepsilon=+1)$
with the potential of Eq.(\ref{Bgp})
%$V\left(  \phi\right)
%=V_{0}\left[  \alpha\cosh\left(  p\phi\right)  +\beta\sinh\left(
%p\phi\right)  \right]  ^{q}$\
and a perfect fluid with equation of state parameter $w_{m}=\left(
\gamma-1\right)  $. Under these assumptions the Lagrangian of the field
equations (\ref{lang.01}) becomes%
\begin{equation}
L\left(  a,\dot{a},\phi,\dot{\phi}\right)  =-3a\dot{a}^{2}+\frac{1}{2}%
a^{3}\dot{\phi}^{2}-V_{0}a^{3}\left[  \alpha\cosh\left(  p\phi\right)
+\beta\sinh\left(  p\phi\right)  \right]  ^{q}-\rho_{m0}a^{-3\left(
\gamma-1\right)  }. \label{ssf.01}%
\end{equation}
From previous work on Noether point symmetries in scalar-field cosmology
\cite{DynSym}, and on dynamical symmetries, \cite{DynSym} we know that this
Lagrangian admits conservation laws when: a. the potential reduces to the
exponential potential i.e. $\beta=\pm\alpha,$ and b. we have the so called
\textit{Unified Dark Matter} (UDM) potential (see Paliathanasis et al.
\cite{DynSym} and references therein), i.e. $p=\frac{\sqrt{6}}{4}$ and $q=2,$
when the extra fluid is dust, namely $(w_{m},\gamma)=(0,1)$.

In the following we consider $\alpha\neq\beta$ which implies that the current
analysis generalizes the previous works of \cite{Barrow,DynSym}. The
Hamiltonian (\ref{lang.02}) of the field equations for the Lagrangian
(\ref{ssf.01}) in terms of the momenta $p_{a},$ and $p_{\phi},$ is
\begin{equation}
-\frac{1}{12a}p_{a}^{2}+\frac{1}{2a^{3}}p_{\phi}^{2}+\left(  V_{0}a^{3}\left[
\alpha\cosh\left(  p\phi\right)  +\beta\sinh\left(  p\phi\right)  \right]
^{q}+\rho_{m0}a^{-3\left(  \gamma-1\right)  }\right)  =0, \label{ssf.02}%
\end{equation}
\ and the WdW equation (\ref{lang.05}) is
\begin{equation}
-\frac{1}{12a}\Psi_{,aa}+\frac{1}{2a^{3}}\Psi_{,\phi\phi}-\frac{1}{12a^{2}%
}\Psi_{,a}+\left(  V_{0}a^{3}\left[  \alpha\cosh\left(  p\phi\right)
+\beta\sinh\left(  p\phi\right)  \right]  ^{q}+\rho_{m0}a^{-3\left(
\gamma-1\right)  }\right)  ~\Psi=0. \label{ssf.03}%
\end{equation}

Applying the results of \cite{AnIJGMMP}, we find that the second-order partial
differential equation (\ref{ssf.03}) admits the generic Lie point symmetry
vector
\begin{equation}
X=\alpha X_{1}+\beta X_{2}+a_{0}\Psi\partial_{\Psi} \label{ssf.04}%
\end{equation}
where $a_{0}$ is a constant\footnote{The fields $X_{1},X_{2}$ are CKVs of the
minisuperspace (\ref{lang.04}).},
\begin{align}
X_{1}  &  =a^{\frac{3\mu}{2}}\left[  \frac{\sqrt{6}}{6}a\sinh\left(
\frac{\sqrt{6}}{4}\mu\phi\right)  \partial_{a}+\cosh\left(  \frac{\sqrt{6}}%
{4}\mu\phi\right)  \partial_{\phi}\right] \label{ssf.05}\\
X_{2}  &  =a^{\frac{3\mu}{2}}\left[  \frac{\sqrt{6}}{6}a\cosh\left(
\frac{\sqrt{6}}{4}\mu\phi\right)  \partial_{a}+\sinh\left(  \frac{\sqrt{6}}%
{4}\mu\phi\right)  \partial_{\phi}\right]  . \label{ssf.06}%
\end{align}
and the constants $p,q,\gamma$ are related as follows:%
\begin{equation}
p=\frac{\sqrt{6}}{4}\mu~,~q=-\frac{4}{\mu}-2~,~\gamma=\mu+2. \label{ssf.07}%
\end{equation}
That is, the effective potential of the field equations is
\begin{equation}
V_{eff}=\left(  V_{0}a^{3}\left[  \alpha\cosh\left(  \frac{\sqrt{6}}{4}\mu
\phi\right)  +\beta\sinh\left(  \frac{\sqrt{6}}{4}\mu\phi\right)  \right]
^{-\frac{4}{\mu}-2}+\rho_{m0}a^{-3\left(  \mu+1\right)  }\right)  .
\label{ssf.08}%
\end{equation}
Therefore, for $\mu=-1$, we have that$~q=2$,~$\gamma=1$ i.e. we have the UDM
potential with dust~(for the exact solution and the observation constraints of
that model see \cite{DynSym}).

If the perfect fluid is a barotropic fluid, that is the barotropic index
$\gamma\in\left[  1,2\right]  $ then, from (\ref{ssf.07}), $\mu\in
\lbrack-1,0)$ since $\mu\neq0$. However, if we require the perfect fluid to
have a negative equation of state parameter, like a cosmological constant,
then $\gamma\in\lbrack0,2)$ which means that $\mu\in\lbrack-2,0)$.
Furthermore, when $\rho_{m0}=0$, i.e. there is no extra fluid, we have that
$\mu\in%
%TCIMACRO{\U{211d} }%
%BeginExpansion
\mathbb{R}
%EndExpansion
^{\ast}$. In the following, we apply the Lie symmetry vector (\ref{ssf.04}) in
order to construct the invariant solution of the WdW equation (\ref{ssf.03})
and to solve the null Hamilton-Jacobi equation \ of the Hamiltonian
(\ref{ssf.02}) in order to reduce the order of the field equations. In the
following section we study the case $\alpha\beta=0$ and in appendix
\ref{general} we present the general solution for $\alpha\beta\neq0$.

\section{Exact solutions for the $\cosh/\sinh$ potential}

\label{ExactSol}

In this section we determine the exact solution of the field equations and of
the WdW equation for the quintessence scalar field. We consider the case
$\alpha=1,~\beta=0$ (the case $\alpha=0,~\beta=1$ is equivalent to that case).
Under the coordinate transformation\footnote{We assume $\mu<0$. However when
$\rho_{m0}=0$ it is possible to have $\mu>0.$ In that case all calculations
remain valid provided we replace $\mu=-\nu$ in (\ref{ssf.08}) and in the
subsequent coordinate transformations.}:
\begin{equation}
a=\left(  x^{2}-y^{2}\right)  ^{-\frac{1}{3\mu}}~,~~\phi=\frac{2\sqrt{6}}%
{3\mu}\arctan h\left(  \frac{y}{x}\right)  , \label{ssf.09}%
\end{equation}
the effective potential (\ref{ssf.08}) becomes%
\begin{equation}
V_{eff}=V_{0}\left(  x^{2}-y^{2}\right)  ^{\frac{1+\mu}{\mu}}x^{-\frac{4}{\mu
}-2} \label{ssf.10}%
\end{equation}
and the WdW equation is
\begin{equation}
\left(  x^{2}-y^{2}\right)  ^{\frac{1}{\mu}+1}\left[  \Psi_{,yy}-\Psi
_{,xx}+\left(  2V_{0}^{\prime}x^{-\frac{4}{\mu}-2}+2\rho_{m0}\right)
\Psi\right]  =0 \label{ssf.11}%
\end{equation}
where $V_{0}^{\prime}=\frac{3}{8}\mu^{2}V_{0},~\rho_{m0}^{\prime}=\frac{3}%
{8}\mu^{2}\rho_{m0}$. In these coordinates the Lie point symmetry vector
(\ref{ssf.04}) is $X=\partial_{y}+a_{0}\Psi\partial_{\Psi}.$ Therefore, the
solution of equation (\ref{ssf.11}) admits an oscillatory term, i.e.$~\Psi
\left(  x,y\right)  =e^{a_{0}y}\Phi\left(  x\right)  $ where%
\begin{equation}
\Phi_{,xx}-\left(  2V_{0}^{\prime}x^{-\frac{4}{\mu}-2}+2\rho_{m0}^{\prime
}+a_{0}^{2}\right)  \Phi=0. \label{ssf.12}%
\end{equation}

Furthermore, in the WKB approximation, $\Psi\propto e^{iS}$ and
%$\Psi\symbol{126}e^{iS}~$and
equation$~$(\ref{ssf.11}) becomes%
\begin{equation}
\left(  x^{2}-y^{2}\right)  ^{\frac{1}{\mu}+1}\left[  \left(  \frac{\partial
S}{\partial y}\right)  ^{2}-\left(  \frac{\partial S}{\partial x}\right)
^{2}+2V_{0}^{\prime}x^{-\frac{4}{\mu}-2}+2\rho_{m0}^{\prime}\right]  =0,
\label{ssf.13}%
\end{equation}
which is the null Hamilton-Jacobi equation which describes the field
equations. The solution of (\ref{ssf.13}) is
\begin{equation}
S\left(  x,y\right)  =c_{1}y\pm\int\sqrt{c_{1}^{2}+2\rho_{m0}^{\prime}%
+2V_{0}^{\prime}x^{-\frac{4}{\mu}-2}}. \label{ssf.14}%
\end{equation}
Therefore, the field equation is reduced to the following two-dimensional
system,
\begin{equation}
\left(  x^{2}-y^{2}\right)  ^{-\left(  \frac{1}{\mu}+1\right)  }\dot{x}%
=\mp\sqrt{c_{1}^{2}+2\rho_{m0}^{\prime}+2V_{0}^{\prime}x^{-\frac{4}{\mu}-2}%
}~,~\left(  x^{2}-y^{2}\right)  ^{-\left(  \frac{1}{\mu}+1\right)  }\dot
{y}=c_{1}. \label{ssf.15}%
\end{equation}

In order to simplify the system (\ref{ssf.15}) further we apply the
transformation $d\tau=$ $\left(  x^{2}-y^{2}\right)  ^{\frac{1}{\mu}%
+1}dt=a^{-3\left(  \mu+1\right)  }dt,$ and the dynamical system becomes
\begin{equation}
x^{\prime}=\mp\sqrt{c_{1}^{2}+2\rho_{m0}^{\prime}+2V_{0}^{\prime}x^{-\frac
{4}{\mu}-2}}~,~y^{\prime}=c_{1} \label{ssf.16}%
\end{equation}
The exact solution of the system (\ref{ssf.16}) is expressed in terms of
elliptic functions\footnote{\textbf{ }It is easy to see that when $\mu=-1$
then $d\tau=dt$, which is the UDM solution for $\omega_{2}=0$ of
\cite{DynSym}.}.

\begin{figure}[ptb]
\includegraphics[height=7cm]{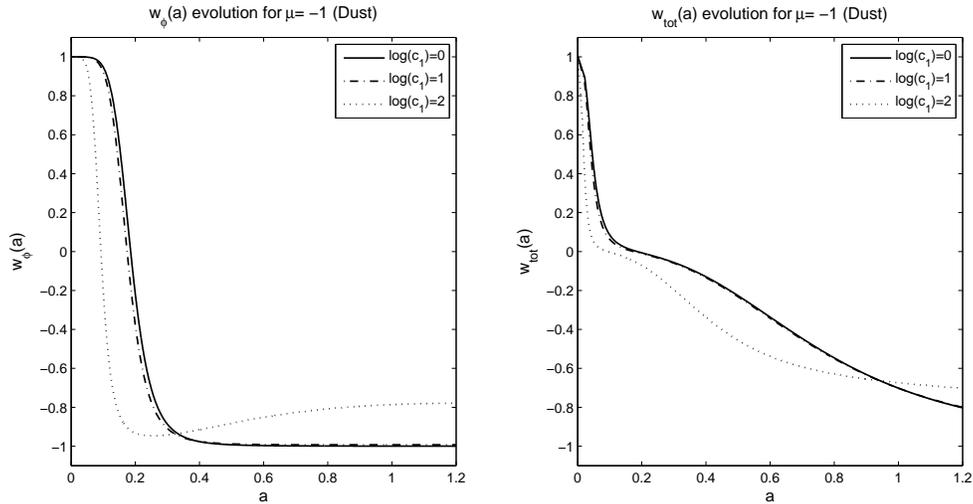}
%\mbox{\epsfxsize=14.2cm \epsffile{numericmodel2q.eps}}
\caption{The evolution of the equation of state parameters $w_{\phi}\left(
a\right)  $ for the scalar field and $w_{tot}\left(  a\right)  $ for the total
fluid for the Lagrangian (\ref{ssf.01}) with $\alpha=1,\beta=0.$ In
(\ref{ssf.15}) we have taken the minus sign and $\mu=-1$ (dust fluid:
$w_{m}=0$). For the numerical solution we use $\left(  x,y\right)
_{t\rightarrow0}=\left(  0.01,0.0099\right)  ,log(V_{0})=4,~$ $\rho
_{m0}=3\Omega_{m0}H_{0}^{2}$ with $\Omega_{m0}=0.308$ and $H_{0}=68$Km/s/Mpc
(or $\rho_{m0}=4.25\times10^{3}$ in units of $8\pi G\equiv1$). The solid line
is for $\mathrm{log}(c_{1})=0$; the dashed line is for $\mathrm{log}%
(c_{1})=1;$ the dotted line is for $\mathrm{log}(c_{1})=2.$}%
\label{fig1dr}%
\end{figure}

We perform a numerical integration of the non-linear system (\ref{ssf.15}) and
in fig. \ref{fig1dr} we give the evolution of the equation of state parameter
for the scalar field $w_{\phi}\left(  a\right)  $ and for the total fluid
$w_{tot}\left(  a\right)  $ for various values of the constant $c_{1}$\ in the
case $\mu=-1$. For $\mu=-1$ the extra perfect fluid is dust, i.e.
$(w_{m},\gamma)=(0,1)$. Concerning the values of the cosmological parameters,
we use the Planck priors \cite{Ade15}, namely $\Omega_{m0}=0.308$ and
$H_{0}=67.8~$km/s/Mpc which imply $\rho_{m0}=3\Omega_{m0}H_{0}^{2}%
\simeq4.25\times10^{3}$, in units of $\kappa=8\pi G\equiv1$. From
fig.\ref{fig1dr}, we observe that the scalar field mimics the cosmological
constant for small values of the constant $c_{1}$, however for large values of
$c_{1}$ the scalar field has an EoS parameter $w_{\phi}>-1$. We find that
within a physical range of the above cosmological parameters the corresponding
dark energy EoS parameter deviates by $\sim1-2\%$. Furthermore, from the
evolution of $w_{tot}(a)$ we see that there is a matter-dominated epoch.
However, as the parameter $c_{1}$ increases, this epoch has shorter duration.
\ In what follows we study the case $c_{1}=0$ and express analytically the
scalar field and the Hubble function in terms of the scale factor. \textbf{ }

\subsection{Subcase $c_{1}=0$.}

\label{sub51}

When $c_{1}=0$, from (\ref{ssf.15}) we have $y\left(  t\right)  =y_{0}$, hence
from (\ref{ssf.09}) it follows that $x^{2}=y_{0}^{2}+a^{-3\mu}$. Furthermore,
from the transformation (\ref{ssf.09}) and from (\ref{ssf.15}) we find that
\begin{equation}
w_{\phi}\left(  a\right)  =\frac{y_{0}^{2}\left[  \frac{\Omega_{m0}}%
{\Omega_{\Lambda0}}+\left(  y_{0}^{2}+a^{-3\mu}\right)  ^{-\frac{2}{\mu}%
-1}\right]  -a^{-3\mu}\left(  y_{0}^{2}+a^{-3\mu}\right)  ^{-\frac{2}{\mu}-1}%
}{y_{0}^{2}\left[  \frac{\Omega_{m0}}{\Omega_{\Lambda0}}+\left(  y_{0}%
^{2}+a^{-3\mu}\right)  ^{-\frac{2}{\mu}-1}\right]  +a^{-3\mu}\left(  y_{0}%
^{2}+a^{-3\mu}\right)  ^{-\frac{2}{\mu}-1}} \label{ssf.19}%
\end{equation}
where we have set
\begin{equation}
\Omega_{\Lambda0}=\frac{V_{0}}{3H_{0}^{2}}~,~\Omega_{m0}=\frac{\rho_{m0}%
}{3H_{0}^{2}}~.
\end{equation}
Here, we would like to note that we call $\Omega_{\Lambda0}~$ the density
parameter of the cosmological constant-like term and $\Omega_{m0}$ the
parameter of the perfect fluid\footnote{In general, for $y_{0}\neq0$,~hold
$\Omega_{m0}+\Omega_{\Lambda0}\neq1$; however the equality holds only when the
constant $y_{0}=0$ [$V(\phi)=V_{0}$] which means that the scalar field act as
a cosmological constant, i.e. $w_{\phi}=-1$, see eq. (\ref{sf.220}).}.

Therefore, for the scalar field density $\rho_{\phi}\left(  a\right)  $ we
have:
\begin{equation}
\rho_{\phi}\left(  a\right)  =3\Omega_{\Lambda0}H_{0}^{2}a^{-6}\left[
y_{0}^{2}\left(  \frac{\Omega_{m0}}{\Omega_{\Lambda0}}+\left(  y_{0}%
^{2}+a^{-3\mu}\right)  ^{-\frac{2}{\mu}-1}\right)  +a^{-3\mu}\left(  y_{0}%
^{2}+a^{-3\mu}\right)  ^{-\frac{2}{\mu}-1}\right]  . \label{ssf.20}%
\end{equation}
Then (\ref{SF.07}) implies that%
\begin{equation}
E^{2}(a)=\frac{H^{2}(a)}{H_{0}^{2}}=\Omega_{m0}a^{-3\left(  \mu+2\right)
}+\Omega_{\Lambda0}a^{-6}\left(  y_{0}^{2}\left[  \frac{\Omega_{m0}}%
{\Omega_{\Lambda0}}+\left(  y_{0}^{2}+a^{-3\mu}\right)  ^{-\frac{2}{\mu}%
-1}\right]  +a^{-3\mu}\left(  y_{0}^{2}+a^{-3\mu}\right)  ^{-\frac{2}{\mu}%
-1}\right)  . \label{ssf.21}%
\end{equation}

We note that if $y_{0}^{2}+a^{-3\mu}\approx a^{-3\mu}$ then%
\begin{equation}
E^{2}(a)=\Omega_{m0}a^{-3\left(  \mu+2\right)  }+\Omega_{\Lambda0}\left(
1+y_{0}^{2}\left[  \frac{\Omega_{m0}}{\Omega_{\Lambda0}}a^{-6}+a^{3\mu
}\right]  \right)  . \label{ssf.22}%
\end{equation}
If $y_{0}=0$, then (\ref{ssf.22}) becomes
%$\frac{H^{2}\left(  a\right)  }{H_{0}^{2}}=\Omega_{m0}a^{-3\left(  \mu+2\right)
$E^{2}(a)=\Omega_{m0}a^{-3\left(  \mu+2\right)  }+\Omega_{\Lambda0}$ which is
obvious because when $y_{0}=0$, we have $\phi=0$ and $V\left(  \phi\right)
=V_{0},$ which means that the scalar field acts as a cosmological constant.
Furthermore, from (\ref{ssf.21}) and $H\left(  a=1\right)  =H_{0}$, we have
the constraint%
\begin{equation}
\left(  1+y_{0}^{2}\right)  \left[  \Omega_{m0}+\Omega_{\Lambda0}\left(
1+y_{0}^{2}\right)  ^{-\frac{2}{\mu}-1}\right]  -1=0. \label{sf.220}%
\end{equation}
In the following section we consider special values of the barotropic constant
$\gamma=\mu+2$.

\subsubsection{Dust fluid versus the effective dark energy EoS}

When the perfect fluid is dust then $\mu=-1,~\gamma=1$ ($w_{m}=0$ for other
cases see appendix A)
%which is the UDM
%potential (without the cosmological constant term) therefore
Eq.(\ref{ssf.21}) takes the following form
\begin{equation}
E^{2}(a)=\Omega_{m0}a^{-3}+\Omega_{\Lambda0}\left[  1+2y_{0}^{2}a^{-3}%
+y_{0}^{2}\left(  \frac{\Omega_{m0}}{\Omega_{\Lambda0}}+y_{0}^{2}\right)
a^{-6}\right]  =\Omega_{m0}a^{-3}+\Delta H(a)\;. \label{sf.22a}%
\end{equation}
It should be mentioned that the last term $\Delta H(a)$ of the normalized
Hubble function (\ref{sf.22a})
%encode
%the $Q$-cosmology corrections to the standard FLRW expression.
%In this case the scalar field
introduces a cosmological constant-like fluid, dust and stiff matter.
Furthermore, from (\ref{sf.220}) we have the following algebraic equation%
\begin{equation}
\Omega_{\Lambda0}y_{0}^{4}+\left(  2\Omega_{\Lambda0}+\Omega_{m_{0}}\right)
y_{0}^{2}+\left(  \Omega_{m0}+\Omega_{\Lambda0}-1\right)  =0 \label{sf.22b}%
\end{equation}
hence, the discriminant of the polynomial (\ref{sf.22b}) (for $y_{0}^{2}$) is%
\begin{equation}
D=\left(  2\Omega_{\Lambda0}+\Omega_{m_{0}}\right)  ^{2}+4\Omega_{\Lambda
0}\left(  1-\Omega_{m0}-\Omega_{\Lambda0}\right)  ,
\end{equation}
and $D\geq0$ when $\left(  1-\Omega_{m0}-\Omega_{\Lambda0}\right)  \geq0$.
Recall that $\Omega_{m0}\in\left[  0,1\right]  ,\Omega_{\Lambda0}\in(0,1],$
and because $y_{0}^{2}>0$ we have the solution
\begin{equation}
y_{0}^{2}=\frac{\sqrt{\left(  \Omega_{m0}\right)  ^{2}+4\Omega_{\Lambda0}%
}-\left(  2\Omega_{\Lambda0}+\Omega_{m0}\right)  }{2\Omega_{\Lambda0}}.
\label{sf.22c}%
\end{equation}
%where $y_{0}^{2}=0$ only when $\Omega_{m0}^{D}=\Omega_{\Lambda0}$.

Let us now compute the effective dark energy EoS $w_{\phi,eff}$ for the scalar
field model introduced above. It is well known that one can express the
effective dark energy EoS parameter in terms of the normalized Hubble
parameter \cite{Saini00}
\begin{equation}
w_{\phi,eff}(a)=\frac{-1-\frac{2}{3}\frac{d\ln E}{d\ln a}}{1-\Omega_{m}(a)}\;,
\label{eos22}%
\end{equation}
where $\Omega_{m}(a)=\frac{\Omega_{m0}a^{-3}}{E^{2}(a)}$ \ . Inserting the
second equality of Eq.(\ref{sf.22a}) into Eq.(\ref{eos22}),
%we find after some simple
%calculations%, one may prove
%that
the effective dark energy EoS parameter takes the following form (see
\cite{Linjen03}):
%given by (see \cite{Linjen03}):%
\begin{equation}
w_{\phi,eff}(a)=-1-\frac{1}{3}\;\frac{d\ln\Delta H}{d\ln a}\; \label{eos222}%
\end{equation}
which implies that any modifications to the effective EoS parameter are
included in the second term of Eq.~(\ref{eos222}). Inserting Eq.~(\ref{sf.22a}%
) into Eq.~(\ref{eos222}) it is straightforward to obtain a simple analytical
expression for the effective dark energy EoS parameter:
\begin{equation}
w_{\phi,eff}(a)=-1+\frac{2y_{0}^{2}a^{-3}+2y_{0}^{2}(\frac{\Omega_{m0}}%
{\Omega_{\Lambda0}}+y_{0}^{2})a^{-6}}{1+2y_{0}^{2}a^{-3}+y_{0}^{2}%
(\frac{\Omega_{m0}}{\Omega_{\Lambda0}}+y_{0}^{2})a^{-6}}.
\end{equation}

\subsubsection{The total case}

\begin{figure}[ptb]
\includegraphics[height=7cm]{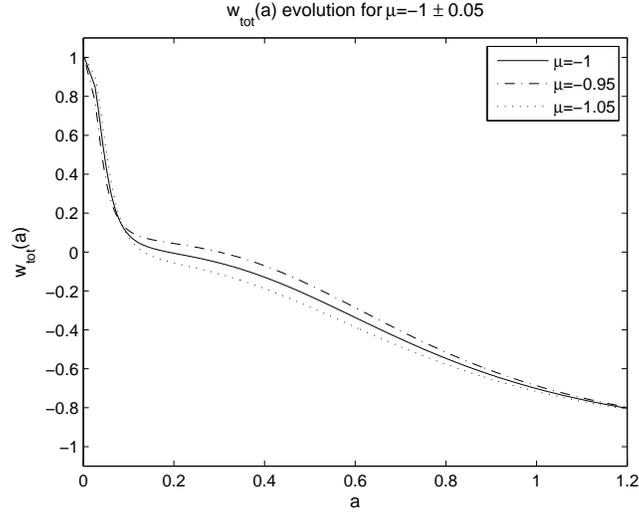}
%\mbox{\epsfxsize=14.2cm \epsffile{dustplusminus.eps}}
\caption{The evolution of the equation of state parameters $w_{tot}\left(
a\right)  $ for the total fluid for the Lagrangian (\ref{ssf.01}) with
$\alpha=1,\beta=0$ where in (\ref{ssf.15}) we take the minus sign and
$c_{1}=0$. ~For the numerical solution we use $\left(  x,y\right)
_{t\rightarrow0}=\left(  0.01,0.0099\right)  ,\mathrm{log}(V_{0})=4~$ and
$\rho_{m0}=4.25\times10^{3}$.$~$The solid line is for $\mu=-1$ [dust:
$(w_{m},\gamma)=(0,1)$], the dash-dotted line is for $\mu=-0.95$ and the
dotted line is for $\mu=-1.05.$}%
\label{fig22}%
\end{figure}

Our dynamical system is integrable in the case of a single perfect fluid.
Here, we introduce two perfect fluids (for example dust and radiation).
%with constant equation of state
%parameter in order to have the value of the effective equation of state
%parameter of two fluids, i.e. dust and radiation fluids.
Consider that we have dust ($w_{m}=0$) and another perfect fluid with equation
of state parameter $P_{f}=w_{f}\rho_{f}$ (for radiation $w_{f}=1/3$), and
$\left(  \frac{\Omega_{f0}}{\Omega_{D0}}\right)  \ll1.$ The latter implies
that the equation of state parameter which is associated with the two perfect
fluids is
\begin{equation}
\bar{w}_{m}=\frac{P_{m}+P_{f}}{\rho_{m}+\rho_{f}}=\frac{w_{f}\rho_{_{f}}}%
{\rho_{m}+\rho_{f}}=\frac{w_{f}\left(  \frac{\Omega_{f}}{\Omega_{m}}\right)
}{1+\left(  \frac{\Omega_{f}}{\Omega_{m}}\right)  }\approx w_{f}\left(
\frac{\Omega_{f}}{\Omega_{m}}\right)  ,~~~\bar{w}_{m}^{2}\approx0\text{ }
\label{ssf.21a}%
\end{equation}
that is, $\gamma=1+\bar{w}_{m}$ and $\mu=-1+\bar{w}_{m}$. When $w_{f}>0$ we
have that $\bar{w}_{m}>0$ and when $w_{f}<0$ holds we have $\bar{w}_{m}<0$. We
replace (\ref{ssf.21a}) in (\ref{ssf.21}) and perform a Taylor expansion near
$\bar{w}_{m}=0$ ($\gamma=1$ or $\mu=-1$). We find%
\begin{equation}
E^{2}(a)=E_{\gamma=1}^{2}(a)-3\Omega_{m0}a^{-3}\ln\left(  a\right)  \bar
{w}_{m}+\Omega_{\Lambda0}F\left(  a\right)  \bar{w}_{m}%
\end{equation}
where the normalized Hubble parameter $E_{\gamma=1}^{2}(a)$ is given by Eq.
(\ref{sf.22a}) and
%$\left(  \frac{H^{2}}{H_{0}^{2}}\right)  _{\left(  \mu=-1\right)  }~$is
%the Hubble function for the UDM solution with dust $\left(  \mu=-1\right)
%,~$i.e. eq. (\ref{sf.22a}), and%
\begin{equation}
F\left(  a\right)  =2\ln\left(  y_{0}^{2}+a^{3}\right)  \left(  1+y_{0}%
^{2}a^{-3}\right)  ^{2}-6\left(  y_{0}^{2}+a^{3}\right)  a^{-3}\ln a\nonumber
\end{equation}
where $F\left(  a\rightarrow1\right)  =2\ln\left(  y_{0}^{2}+1\right)  \left(
1+y_{0}^{2}\right)  ^{2}$ and when $y_{0}=0$, $F\left(  a\right)  =0$. In fig.
(\ref{fig22}) we give the numerical solutions of the total EoS parameter
$w_{tot}\left(  a\right)  $ for $\mu=1\pm0.05$ and $c_{1}=0$.

In appendix \ref{appenA}, we give the exact solutions for the Hubble function,
$H\left(  a\right)  $, for the cases where $\gamma=\frac{4}{3}$ (radiation
fluid) and $\gamma=\frac{2}{3}$ (curvature-like fluid).

\section{Dynamical Analysis}

\label{dynanalysis} In order to complete our analysis of the model with
Lagrangian (\ref{ssf.01}), we perform a dynamical analysis of the field
equations by studying the fixed points of the field equations. We introduce
the new dimensionless variables \cite{Ame10,copeland}%
\begin{equation}
x=\frac{\dot{\phi}}{\sqrt{6}H}~,~y=\frac{\sqrt{V}}{\sqrt{3}H}~,~\Omega
_{m}=\frac{\rho_{m}}{3H^{2}}~,~\lambda=-\frac{V_{,\phi}}{V} \label{ssf.47}%
\end{equation}
and the lapse time $N=\ln a$. In the new variables the field equations reduce
to the following first-order ODEs
\begin{equation}
\frac{dx}{dN}=-3x+\frac{\sqrt{6}}{2}\lambda y^{2}+\frac{3}{2}x\left[  \left(
1-w_{m}\right)  x^{2}+\left(  1+w_{m}\right)  \left(  1-y^{2}\right)  \right]
\label{ssf.48}%
\end{equation}%
\begin{equation}
\frac{dy}{dN}=-\frac{\sqrt{6}}{2}\lambda xy+\frac{3}{2}y\left[  \left(
1-w_{m}\right)  x^{2}+\left(  1+w_{m}\right)  \left(  1-y^{2}\right)  \right]
\label{ssf.49}%
\end{equation}%
\begin{equation}
\frac{d\lambda}{dN}=-\sqrt{6}\lambda^{2}\left(  \Gamma-1\right)  x
\label{ssf.50}%
\end{equation}
where~$\Gamma=\frac{V_{,\phi\phi}V}{V_{,\phi}^{2}}$ and the Friedmann equation
(\ref{SF.07}) gives the constraint $\Omega_{m}=1-\Omega_{\phi}$, where
$\Omega_{\phi}=x^{2}+y^{2}$.

In this case the second Friedmann equation (\ref{frie2}) becomes%
\begin{equation}
\frac{2}{3}\frac{\dot{H}}{H^{2}}=-1-w_{m}-\left(  1-w_{m}\right)
x^{2}-\left(  1+w_{m}\right)  \left(  1-y^{2}\right)  \label{ssf.51}%
\end{equation}
which gives that the total EoS parameter $w_{tot}$ as a function of $w_{m},$
$x~$and $y$:%

\begin{equation}
w_{tot}=w_{m}+\left(  1-w_{m}\right)  x^{2}-\left(  1+w_{m}\right)  y^{2}.
\label{ssf.52}%
\end{equation}
Furthermore, the EoS parameter $w_{\phi}~$for the scalar field is $w_{\phi
}=\frac{x^{2}-y^{2}}{x^{2}+y^{2}}$. Note that at any point $\left(
x_{0},y_{0},\lambda\right)  ,$ from (\ref{ssf.51}) the solution of the scalar
factor is a power law as long as $w_{tot}=const.;$ that is,
%$a\left(t\right)  =a_{0}\left(  t-t_{0}\right)  ^{\frac{2}{3\left(  1+w_{tot}\right)}}$
$a\left(  t\right)  \propto t^{\frac{2}{3\left(  1+w_{tot}\right)  }}$ for
$w_{tot}\neq-1$ and $a\left(  t\right)  =a_{0}e^{H_{0}t}$ for $w_{tot}=-1$ $.$

In the following we consider in (\ref{ssf.01}) $\beta=0,$ so that the
potential of the scalar field is $V\left(  \phi\right)  =V_{0}\cosh^{q}\left(
p\phi\right)  $ \cite{cop2,pavl,Fadragas}. For this potential we write
$\Gamma\left(  \phi\right)  $ as a function of $\lambda$, i.e. $\Gamma\left(
\lambda\right)  =1+\frac{qp^{2}}{\lambda^{2}}-\frac{1}{q}$, and equation
(\ref{ssf.50}) becomes%

\begin{equation}
\frac{d\lambda}{dN}=-\frac{\sqrt{6}}{q}\left(  qp-\lambda\right)  \left(
qp+\lambda\right)  x. \label{ssf.53}%
\end{equation}

Equations (\ref{ssf.48}), (\ref{ssf.49}) and (\ref{ssf.53}) describe an
autonomous dynamical system in the $E^{3}$ space. Furthermore from the
constraints $0\leq\Omega_{\phi}\leq1,~y\geq0,$ the variables $\left(
x,y\right)  $ are bounded in the ranges $x\in\left[  -1,1\right]  $%
,~$y\in\left[  0,1\right]  ~$from which follows that that the points $\left(
x,y\right)  $ belong to a half disk; however for the parameter $\lambda$ there
is no constraint that implies that $\lambda\in%
%TCIMACRO{\U{211d} }%
%BeginExpansion
\mathbb{R}
%EndExpansion
$ \cite{copeland,Tamanini}.
%If the perfect fluid $\rho_{m}$ is a baryonic fluid then
%$w_{m}\in\left[  0,1\right]$.
Furthermore, we consider $w_{m}\in\left(  -1,1\right)  $. The fixed points of
the dynamical system (\ref{ssf.48}), (\ref{ssf.49}) and (\ref{ssf.53}) and the
corresponding cosmological parameters are given in Table \ref{fixedpoints}.
The eigenvalues of the linearized dynamical system near the fixed points are
given in table \ref{eigenvalues}\footnote{Where in table \ref{eigenvalues}
$\Delta=\sqrt{\left(  1-w_{m}\right)  \left(  24\left(  1+w_{m}\right)
^{2}-\left(  7+9w_{m}\right)  \left(  qp\right)  ^{2}\right)  }$}.%

%TCIMACRO{\TeXButton{B}{\begin{table}[tbp] \centering}}%
%BeginExpansion
\begin{table}[tbp] \centering
%EndExpansion
\caption{Fixed points and cosmological parameters}%
\begin{tabular}
[c]{cccccc}\hline\hline
\textbf{Point} & $\left(  \mathbf{x,y,\lambda}\right)  $ & $\mathbf{\Omega
}_{m}$ & $\mathbf{w}_{tot}$ & $\mathbf{w}_{\phi}$ & Acceleration\\\hline
$O$ & $\left(  0,0,\lambda\right)  $ & $1$ & $w_{m}$ & $\nexists$ &
$w_{m}<-\frac{1}{3}$\\
$A_{\left(  \pm\right)  }$ & $\left(  1,0,\pm qp\right)  $ & $0$ & $1$ & $1$ &
No\\
$B_{\left(  \pm\right)  }$ & $\left(  -1,0,\pm qp\right)  $ & $0$ & $1$ & $1$
& No\\
$C$ & $\left(  0,1,0\right)  $ & $0$ & $-1$ & $-1$ & Yes\\
$D_{\left(  \pm\right)  }$ & $\left(  \pm\frac{\sqrt{6}}{6}qp,\sqrt
{1-\frac{\left(  qp\right)  ^{2}}{6}},\pm qp\right)  $ & $0$ & \thinspace
$-1+\frac{\left(  qp\right)  ^{2}}{3}$ & $-1+\frac{\left(  qp\right)  ^{2}}%
{3}$ & $\left\vert qp\right\vert <\sqrt{2}$\\
$E_{\left(  \pm+\right)  }$ & $\left(  \pm\frac{\sqrt{6}\left(  1+w_{m}%
\right)  }{2qp},\frac{\sqrt{6}\sqrt{1-w_{m}^{2}}}{2qp},\pm qp\right)  $ &
$1-\frac{3\left(  1+w_{m}\right)  }{\left(  qp\right)  ^{2}}$ & $w_{m}$ &
$w_{m}$ & $w_{m}<-\frac{1}{3}$\\
$E_{\left(  \pm-\right)  }$ & $\left(  \pm\frac{\sqrt{6}\left(  1+w_{m}%
\right)  }{2qp},-\frac{\sqrt{6}\sqrt{1-w_{m}^{2}}}{2qp},\pm qp\right)  $ &
$1-\frac{3\left(  1+w_{m}\right)  }{\left(  qp\right)  ^{2}}$ & $w_{m}$ &
$w_{m}$ & $w_{m}<-\frac{1}{3}$\\\hline\hline
\end{tabular}
\label{fixedpoints}%
%TCIMACRO{\TeXButton{E}{\end{table}}}%
%BeginExpansion
\end{table}%
%EndExpansion
%

%TCIMACRO{\TeXButton{B}{\begin{table}[tbp] \centering}}%
%BeginExpansion
\begin{table}[tbp] \centering
%EndExpansion
\caption{Eigenvalues of fixed points}%
\begin{tabular}
[c]{cccc}\hline\hline
\textbf{Point} & $\mathbf{m}_{1}$ & $\mathbf{m}_{2}$ & $\mathbf{m}_{3}%
$\\\hline
$O$ & $0$ & $\frac{3}{2}\left(  1+w_{m}\right)  $ & $-\frac{3}{2}\left(
1-w_{m}\right)  $\\
$A_{\left(  \pm\right)  }$ & $3\left(  1-w_{m}\right)  $ & $\pm2p\sqrt{6}$ &
$3\mp\frac{\sqrt{6}}{2}qp$\\
$B_{\left(  \pm\right)  }$ & $3\left(  1-w_{m}\right)  $ & $\mp2p\sqrt{6}$ &
$3\pm\frac{\sqrt{6}}{2}qp$\\
$C$ & $-3\left(  1+w_{m}\right)  $ & $-\frac{3}{2}\left(  1-\sqrt{1-4qp^{2}%
}\right)  $ & $-\frac{3}{2}\left(  1+\sqrt{1-4qp^{2}}\right)  $\\
$D_{\left(  \pm\right)  }$ & $-3\left(  1+w_{m}\right)  +\left(  qp\right)
^{2}$ & $-3+\frac{\left(  qp\right)  ^{2}}{2}$ & $2qp^{2}$\\
$E_{\left(  \pm,\pm\right)  }$ & $\frac{6}{q}\left(  1+w_{m}\right)  $ &
$-\frac{3}{4}\left[  \left(  1-w_{m}\right)  +\frac{\Delta}{qp}\right]  $ &
$-\frac{3}{4}\left[  \left(  1-w_{m}\right)  -\frac{\Delta}{qp}\right]
$\\\hline\hline
\end{tabular}
\label{eigenvalues}%
%TCIMACRO{\TeXButton{E}{\end{table}}}%
%BeginExpansion
\end{table}%
%EndExpansion

Point$~O$ exists for all values of the parameter $\lambda$ and corresponds to
the matter epoch $\left(  \Omega_{m}=1\right)  ;$ the total EoS parameter is
$w_{tot}=w_{m}$ . Since there exists at least one positive eigenvalue,
$m_{2}>0,$ the point $O$ is always unstable. At this point the universe
accelerates if and only if $w_{m}<-\frac{1}{3}$. At the points $A_{\left(
\pm\right)  }$ and $B_{\left(  \pm\right)  }$ the universe is dominated by the
kinetic energy of the scalar field $\left(  \Omega_{m}=1,~V\left(
\phi\right)  =0\right)  $ which means that the scalar field acts as a stiff
fluid, i.e. $\rho_{\phi}\propto a^{-6}~$ which provides a decelerating
universe. These points exist when $\lambda=\pm qp,$ for arbitrary $q,p$. For
these points there exist positive eigenvalues of the linearized system,
$m_{1}>0$, for $w_{m}\in\left(  -1,1\right)  $, hence these critical points
are always unstable.

Point $C$ is the de Sitter solution, $\left(  \Omega_{m}=0\text{,}%
~w_{tot}=-1\right)  $ where the scalar field acts as a cosmological constant
and the matter component vanished. This point exists for all values of the
constants $q,p$ and could be the future attractor of the universe. From the
eigenvalues of Table \ref{eigenvalues} for that point we have that it is
stable when $q>0$ (for a similar solution see \cite{HEW}). The points
$D_{\left(  \pm\right)  }$ correspond to a scalar field dominated
universe~$\left(  \Omega_{m}=0\right)  $ and exist only when $\left\vert
qp\right\vert <\sqrt{6}.$ The total EoS parameter is that of the scalar field
$w_{\phi}=-1+\frac{\left(  qp\right)  ^{2}}{3}$ which gives an accelerated
universe when $|qp|<\sqrt{2}$. The points $D_{\left(  \pm\right)  }$ are
stable for $q<0$ and$~\left\vert qp\right\vert <\sqrt{3\left(  1+w_{m}\right)
}$. Hence, we see that $D_{\left(  \pm\right)  }$ are stable points and
describe an accelerated universe when $w_{\phi}<-\frac{1}{3}$ and
$|qp|<\sqrt{2}$. \ Furthermore, in the limit $q\rightarrow0^{-},$ these points
correspond to the de Sitter universe; when $q\rightarrow0^{-}$ then $V\left(
\phi\right)  \rightarrow V_{0}$.

Finally, the points $E_{\left(  \pm,\pm\right)  }$ are the so called 'scaling'
solutions where $\Omega_{m}=1-\frac{3\left(  1+w_{m}\right)  }{\left(
qp\right)  ^{2}}$ and the scalar field mimics the matter component of the
universe, i.e. $w_{\phi}=w_{m}.$ The points $E_{\left(  \pm,+\right)  }~$exist
when $qp>\sqrt{3\left(  1+w_{m}\right)  }~$\ and they are stable when $q,p<0$
whereas the points $E_{\left(  \pm,-\right)  }$ exist when $qp<-\sqrt{3\left(
1+w_{m}\right)  }$ and are stable when $q<0$, $p>0$. The total EoS parameter
is $w_{tot}=w_{m}$ so they lead to an accelerated universe when $w_{m}%
<-\frac{1}{3}$.

\begin{figure}[ptb]
\includegraphics[height=10cm]{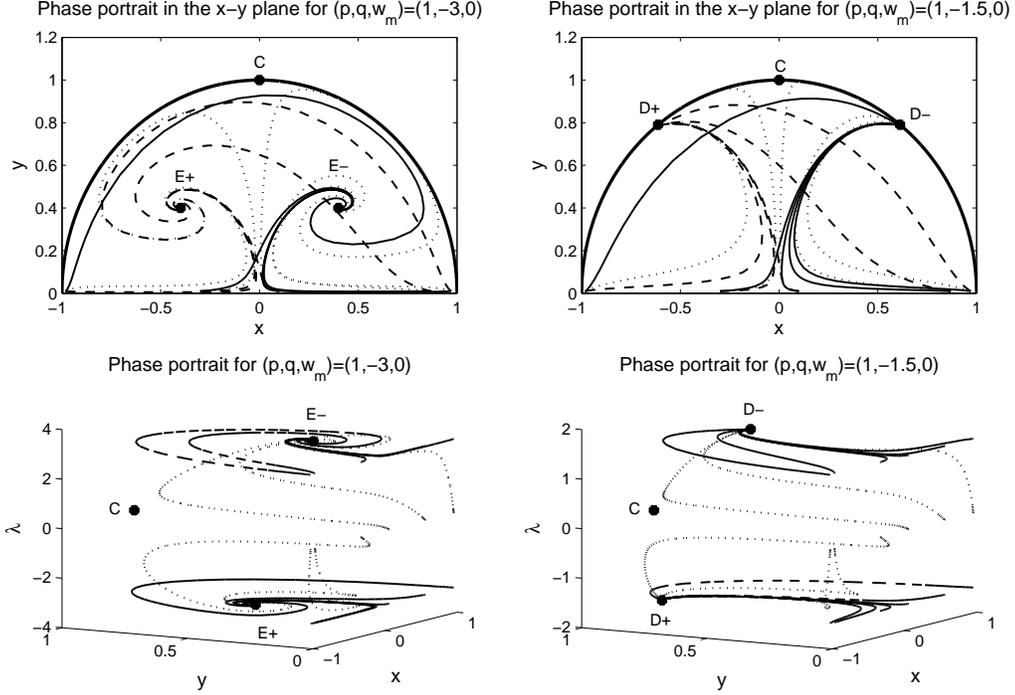}
%\mbox{\epsfxsize=14.2cm \epsffile{numericmodel2q.eps}}
\caption{Phase portrait in the $x-y$ plane and in the $E^{3}$ space for the
potential $V\left(  \phi\right)  =V_{0}\cosh^{q}\left(  p\phi\right)  .~$
Left-hand figures are for the variables $\left(  p,q,w_{m}\right)  =\left(
1,-1.5,0\right)  $ and the right-hand figures are for the variables $\left(
p,q,w_{m}\right)  =\left(  1,-3,0\right)  .$ For $(pq)^{2}>{3,}~$the stable
points are the points $E_{(\pm)}$ (scaling solutions) while for $(pq)^{2}<3$
the two stable points are the points $D_{\pm}$. The solid lines are for the
initial condition $\lambda=pq$, the dashed lines for $\lambda=-pq,$ and the
dotted lines for $\lambda=0$. }%
\label{dyne}%
\end{figure}

In fig. \ref{dyne}, we give the two-dimensional phase portrait in the $x-y$
plane and the three-dimensional phase portrait of the model with values
$\left(  p,q,w_{m}\right)  =\left(  1,-1.5,0\right)  $ and $\left(
p,q,w_{m}\right)  =\left(  1,-3,0\right)  $. We observe that for $(pq)^{2}<3$
the two stable points are the points $D_{\left(  \pm\right)  }$ whereas for
$(pq)^{2}>{3}$ the stable points are $E_{(\pm,-)}$.

It is important to study the case when the constants $q,p,w_{m}$ $\ $are
related to the constant $\mu$ so as to render the integrable field equations
(\ref{ssf.07}). This is the case we studied in the previous section. Since we
considered$~w_{m}\in\left(  -1,1\right)  $ we have that $\mu\in\left(
-2,0\right)  $.

Hence, for the integrable case we have that the points $O,~A_{\left(
\pm\right)  },$ $B_{\left(  \pm\right)  },$ $D_{\left(  \pm\right)  }~$exist
and they are always unstable. The point $O$ has $w_{tot}<-\frac{1}{3}$ when
$\mu\in\left(  -2,-\frac{4}{3}\right)  $ and the points $D_{\left(
\pm\right)  }$ describe an accelerated universe so long as $\mu\in\left(
-2,-2+\frac{2\sqrt{3}}{3}\right)  $. The point $C$ exists and it is the unique
stable point. Finally, the tracker solutions, i.e. points $E_{\left(  \pm
,\pm\right)  },$ do not exist for $\mu\in\left(  -2,0\right)  $. The existence
and the stability of the fixed points for general values $q,p$ and for the
integrable case are given in Table \ref{stability}.%

%TCIMACRO{\TeXButton{B}{\begin{table}[tbp] \centering}}%
%BeginExpansion
\begin{table}[tbp] \centering
%EndExpansion
\caption{Fixed points and their stability for the  general potential and for the integrable subcases}%
\begin{tabular}
[c]{ccccc}\hline\hline
\textbf{Point} & \textbf{Existence} & \textbf{Stability} & \textbf{Stability
for }$\mathbf{\mu}\in\left(  -2,0\right)  $ & \textbf{Acceleration}\\\hline
$O$ & $p,q\in%
%TCIMACRO{\U{211d} }%
%BeginExpansion
\mathbb{R}
%EndExpansion
^{\ast}$ & Unstable & Unstable & $\mu\in\left(  -2,-\frac{4}{3}\right)  $\\
$A_{\left(  \pm\right)  }$ & $p,q\in%
%TCIMACRO{\U{211d} }%
%BeginExpansion
\mathbb{R}
%EndExpansion
^{\ast}$ & Unstable & Unstable & No\\
$B_{\left(  \pm\right)  }$ & $p,q\in%
%TCIMACRO{\U{211d} }%
%BeginExpansion
\mathbb{R}
%EndExpansion
^{\ast}$ & Unstable & Unstable & No\\
$C$ & $p,q\in%
%TCIMACRO{\U{211d} }%
%BeginExpansion
\mathbb{R}
%EndExpansion
^{\ast}$ & Stable for $q\in%
%TCIMACRO{\U{211d} }%
%BeginExpansion
\mathbb{R}
%EndExpansion
^{\ast+}$ & Stable & Yes\\
$D_{\left(  \pm\right)  }$ & $\left\vert qp\right\vert <\sqrt{6}$ & Stable for
$q\in%
%TCIMACRO{\U{211d} }%
%BeginExpansion
\mathbb{R}
%EndExpansion
^{\ast-}~,~\left\vert qp\right\vert <\sqrt{3\left(  1+w_{m}\right)  }$ &
Unstable & $\mu\in\left(  -2,-2+\frac{2\sqrt{3}}{3}\right)  $\\
$E_{\left(  \pm,+\right)  }$ & $qp>\sqrt{3\left(  1+w_{m}\right)  }$ &
Stable~for~$q\in%
%TCIMACRO{\U{211d} }%
%BeginExpansion
\mathbb{R}
%EndExpansion
^{\ast-}~$,~$p<0$ & $\nexists$ & $\nexists$\\
$E_{\left(  \pm,-\right)  }$ & $qp<-\sqrt{3\left(  1+w_{m}\right)  }$ & Stable
for~$q\in%
%TCIMACRO{\U{211d} }%
%BeginExpansion
\mathbb{R}
%EndExpansion
^{\ast-}~,~p>0$ & $\nexists$ & $\nexists$\\\hline\hline
\end{tabular}
\label{stability}%
%TCIMACRO{\TeXButton{E}{\end{table}}}%
%BeginExpansion
\end{table}%
%EndExpansion

\begin{figure}[ptb]
\includegraphics[height=7cm]{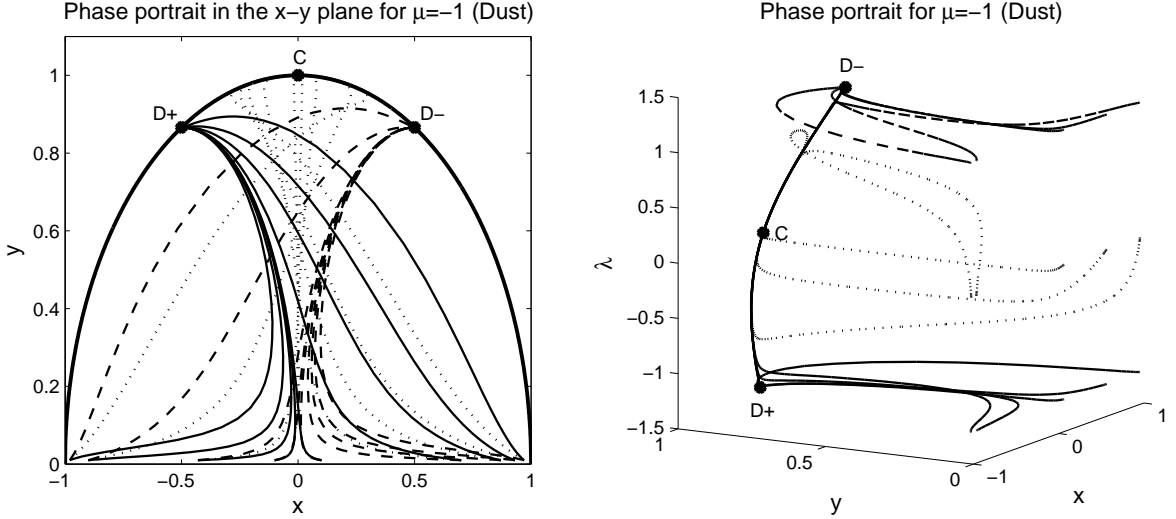}
%\mbox{\epsfxsize=14.2cm \epsffile{numericmodel2q.eps}}
\caption{Phase portrait in the $x-y$ plane and in the $E^{3}$ space for the
potential $V\left(  \phi\right)  =V_{0}\cosh^{q}\left(  p\phi\right)  $ with
variables $\left(  p,q,w_{m}\right)  =\left(  -\frac{\sqrt{6}}{4},2,0\right)
,$ which corresponds to the integrable case for $\mu=-1$. The point $C$ is the
unique stable point. \ The points $D_{\left(  \pm\right)  }$ act as attractors
in the plane $\left(  x-y\right)  $ for $\lambda=\pm\frac{\sqrt{6}}{2}$. From
the right-hand plot we observe that the solutions reach a boundary where
$\Omega_{m0}=0,$ $w_{\phi}>-1$ and from there they move to the de Sitter
point~$C$. \ The solid lines are for initial condition $\lambda=pq$, the
dashed lines for $\lambda=-pq,$ and the dotted lines for $\lambda=0$. }%
\label{dyndust}%
\end{figure}

In fig. \ref{dyndust} we give the two-dimensional phase portrait in the $x-y$
plane and the three-dimensional phase portrait of the the model with values
$\left(  p,q,w_{m}\right)  =\left(  -\frac{\sqrt{6}}{4},2,0\right)  ,$ which
correspond to the integrable case for $\mu=-1$. The point $C$ is the unique
stable point. We observe that the points $D_{\left(  \pm\right)  }$ act as
attractors in the plane $\left(  x-y\right)  $ for $\lambda=\pm\frac{\sqrt{6}%
}{2}$ and the solutions\ reach the boundary where $\Omega_{m}=0$, and move to
the de\ Sitter points $\left(  w_{\phi}=-1\right)  $. It is important to note
that the existence of conservation laws in the field equations which follow
from the Lie point symmetries of the WdW equation, i.e. the dynamical system
is Liouville integrable, gives us constraints on the free parameters of the
model so that there exists a unique stable point which describes the de Sitter universe.

\section{CONCLUSIONS}

\label{conclusion}

We have applied Lie symmetry methods in order to extend the works of Rubano
and Barrow \cite{Barrow} and Paliathanasis et al. \cite{DynSym} in scalar
field cosmology for a general family of potentials, $V(\phi)$. We have shown
that there exists a unique connection between the Lie point symmetries of the
WdW equation and the conservation laws of the field equations. We considered a
general form of $V(\phi)$ which contains hyperbolic functions for a scalar
field with a perfect fluid and we have investigated the existence of Lie point
symmetries of the WdW equation. This approach is more general than the
application of Noether point symmetries.
%We have found that the WDW equation admits Lie
%symmetries and thus invariant solutions if the free parameters are related.
We recovered the result of \cite{Barrow}, that is, that in scalar field
cosmology amongst the variety of $V(\phi)$ potentials the hyperbolic types
play a key role because they admit conservation laws. Moreover, based on the
Lie point symmetries of the WdW equation, we have obtained the exact solutions
of the field equations. Finally, we have performed a dynamical system analysis
by studying the fixed points of the field equations in dimensionless
variables. We found various dynamical cases among which, if the current
cosmological model is Liouville integrable there is a unique stable point
which describes the de-Sitter universe as a late-time attractor for the dynamics.

\begin{acknowledgments}
AP acknowledges financial support of INFN. JDB acknowledges support from the STFC.
\end{acknowledgments}

\appendix

\section{Solutions with radiation and curvature}

\label{appenA}

Here we provide some details concerning the solutions of section \ref{sub51}.
Specifically, for a radiation perfect fluid, $\gamma=\frac{4}{3}$ (hence
$\mu=-\frac{2}{3}$ ($w_{m}=1/3$) and $\Omega_{r0}\equiv\Omega_{m0})$
Eq.(\ref{ssf.21}) gives
\begin{equation}
E^{2}\left(  a\right)  =\frac{H^{2}\left(  a\right)  }{H^{2}_{0}} =\Omega
_{r0}a^{-4}+\Omega_{\Lambda0}\left[  1+3y_{0}^{2}a^{-2}+3y_{0}^{4}a^{-4}%
+y_{0}^{2}\left(  \frac{\Omega_{r0}}{\Omega_{\Lambda0}}+y_{0}^{2}\right)
a^{-6}\right]  \;.
\end{equation}
%which means that the scalar field introduces a cosmological constant-like
%fluid, a curvature like fluid, radiation and stiff matter

%\subsubsection{Curvature-like fluid:}
On the other hand, when $~\gamma=\frac{2}{3}$ (or $\mu=-\frac{4}{3}$), the
perfect fluid has an equation of state $w_{m}=-\frac{1}{3}$ (which can also be
seen as the curvature term in a non-flat FLRW spacetime); then $\Omega
_{K0}\equiv\Omega_{m0}$ and:%
\begin{equation}
E^{2}\left(  a\right)  =\frac{H^{2}\left(  a\right)  }{H^{2}_{0}}=\Omega
_{K0}a^{-2}+\Omega_{\Lambda0}\left[  \sqrt{y_{0}^{2}+a^{4}}a^{-2}+y_{0}%
^{2}\left(  \frac{\Omega_{K0}}{\Omega_{\Lambda0}}+\sqrt{y_{0}^{2}+a^{4}%
}\right)  a^{-6}\right]  .
\end{equation}
This is a solution of the scalar field cosmology in a curved FLRW spacetime.

\section{Exact solution for a general potential}

\label{general} In the general case where $\alpha\neq0$ and $\beta=1$ we apply
the following coordinate transformations,%
\begin{equation}
a=\left[  \left(  x-\frac{y}{\alpha}\right)  ^{2}-y^{2}\right]  ^{-\frac
{1}{3\mu}}~,~\phi=\frac{2\sqrt{6}}{3\mu}\arctan h\left(  \frac{y}{x-\frac
{y}{\alpha}}\right)  .
\end{equation}

In the new coordinate system the WdW equation becomes%
\begin{equation}
\left[  \left(  x-\frac{y}{\alpha}\right)  ^{2}-y^{2}\right]  ^{\frac{1+\mu
}{\mu}}\left[  -\left(  1-\frac{1}{\alpha^{2}}\right)  \Psi_{,xx}+\frac
{2}{\alpha}\Psi_{,xy}+\Psi_{,yy}+\left(  2V_{0}^{\prime}\alpha^{-\frac{4}{\mu
}-2}x^{-\frac{4}{\mu}-2}+2\rho_{m0}^{\prime}\right)  \Psi\right]  =0,
\label{ssf.281}%
\end{equation}
where $V_{0}^{\prime}=\frac{3}{8}\mu^{2}\alpha^{-\frac{4}{\mu}-2}V_{0}%
,~\rho_{m0}^{\prime}=\frac{3}{8}\mu^{2}\rho_{m0}$ and the Lie
%,~\rho_{m0}^{\prime}=\frac{3}{8}\mu^{2}\rho_{m0}^{\prime}$ and the Lie
symmetry vector\ is (\ref{ssf.04}) $X=\partial_{y}+a_{0}\Psi\partial_{\Psi}.$
Therefore, the invariant solution of the WdW equation (\ref{ssf.281}) is
$\Psi\left(  x,y\right)  =e^{a_{0}y}\bar{\Phi}\left(  x\right)  $ where
$\bar{\Phi}\left(  x\right)  $ satisfies the following second-order ODE%
\begin{equation}
\left[  -\left(  1-\frac{1}{\alpha^{2}}\right)  \Phi_{,xx}+\frac{2\alpha_{0}%
}{\alpha}\Phi_{,x}+\left(  2V_{0}^{\prime}x^{-\frac{4}{\mu}-2}+2\rho
_{m0}^{\prime}+\alpha_{0}^{2}\right)  \Phi\right]  =0. \label{ssf.2821}%
\end{equation}
When $\alpha=1,$ which is the case of the exponential scalar field, from
(\ref{ssf.2821}) we have
\begin{equation}
\bar{\Phi}\left(  x\right)  =\exp\left(  -\frac{\left(  2\rho_{m0}^{\prime
}+a_{0}^{2}\right)  }{2a_{0}}x+\frac{V_{0}^{\prime}}{a_{0}\left(  \frac{4}%
{\mu}+1\right)  }x^{-\frac{4}{\mu}-1}\right)  .
\end{equation}

Furthermore, in the WKB approximation the WdW equation (\ref{ssf.281}) becomes
the null Hamilton-Jacobi equation%
\begin{equation}
\left[  \left(  x-\frac{y}{\alpha}\right)  ^{2}-y^{2}\right]  ^{\frac{1+\mu
}{\mu}}\left[  -\left(  1-\frac{1}{a^{2}}\right)  \left(  \frac{\partial
\bar{S}}{\partial x}\right)  ^{2}+\frac{2}{\alpha}\left(  \frac{\partial
\bar{S}}{\partial x}\right)  \left(  \frac{\partial\bar{S}}{\partial
y}\right)  ^{2}+\left(  \frac{\partial\bar{S}}{\partial y}\right)
^{2}+\left(  2V_{0}^{\prime}x^{-\frac{4}{\mu}-2}+2\rho_{m0}^{\prime}\right)
\right]  =0 \label{sf.021}%
\end{equation}
with solution of the form $\ \bar{S}\left(  x,y\right)  =S_{1}\left(
x\right)  +c_{1}y,~$where%
\begin{equation}
S_{1}\left(  x\right)  =\mp\frac{\alpha}{1-\alpha^{2}}\int\left(  c_{1}%
+\sqrt{2V_{0}^{\prime}\left(  \alpha^{2}-1\right)  x^{-\frac{4}{\mu}-2}%
+\alpha^{2}c_{1}^{2}+2\left(  a^{2}-1\right)  \rho_{m0}^{\prime}}\right)
~,~\left\vert \alpha\right\vert \neq1 \label{ssf.32a}%
\end{equation}
and%
\begin{equation}
S_{1}\left(  x\right)  =\int\left(  -\frac{V_{0}^{\prime}}{c_{1}}x^{-\frac
{4}{\mu}-2}-\frac{1}{2}c_{1}-\frac{\rho_{m0}^{\prime}}{c_{1}}\right)
dx~~,~\alpha=1.~
\end{equation}

From the solution of the Hamilton-Jacobi equation (\ref{sf.021}) we can reduce
the equivalent Hamiltonian system of the field equation to the following
system of first order equations%
\begin{align}
\dot{x}  &  =\left[  -\left(  1-\frac{1}{\alpha^{2}}\right)  p_{x}+\frac
{1}{\alpha}p_{y}\right]  \left[  \left(  x-\frac{y}{\alpha}\right)  ^{2}%
-y^{2}\right]  ^{\frac{1+\mu}{\mu}}~,\label{ssf.33}\\
\dot{y}  &  =\left[  \frac{1}{\alpha}p_{x}+p_{y}\right]  \left[  \left(
x-\frac{y}{\alpha}\right)  ^{2}-y^{2}\right]  ^{\frac{1+\mu}{\mu}}~,
\label{ssf.34}%
\end{align}
where $p_{x}=\frac{\partial\bar{S}}{\partial x}$ and $p_{y}=\frac{\partial
\bar{S}}{\partial y}$. \ 

In order to make the reduced system simpler, we apply the transformation
$d\tau=a^{-3\left(  \mu+1\right)  }dt,~$and the system (\ref{ssf.33}%
)-(\ref{ssf.34}) becomes%
\begin{equation}
x^{\prime}=-\left(  1-\frac{1}{\alpha^{2}}\right)  p_{x}+\frac{1}{\alpha}%
p_{y}~~,~~y^{\prime}=\frac{1}{\alpha}p_{x}+p_{y}. \label{ssf.37}%
\end{equation}
For the exponential potential~$\left(  \alpha=1\right)  $, from system
(\ref{ssf.37}) we find the solution in closed form. The solution is
\begin{equation}
x\left(  \tau\right)  =c_{1}\tau+c_{0},
\end{equation}%
\begin{equation}
y\left(  \tau\right)  =\frac{V_{0}^{\prime}}{c_{1}^{2}\left(  \frac{4}{\mu
}+1\right)  }\left(  c_{1}\tau+c_{0}\right)  ^{-\frac{4}{\mu}-1}-\frac
{\rho_{m0}^{\prime}}{c_{1}}\tau+\frac{c_{1}}{2}\tau+y_{0},
\end{equation}
where the scale factor is $a\left(  \tau\right)  =\left(  x^{2}\left(
\tau\right)  -2x\left(  \tau\right)  y\left(  \tau\right)  \right)
^{-\frac{1}{3\mu}}.$ This is the solution of the exponential scalar field with
matter in the Einstein frame. Recall that the matter has an equation of state
parameter of the form $w_{m}=\mu+1$.

\subsection{Special solution with $\rho_{m0}=0$, $c_{1}=0$}

When $\rho_{m0}=0$ and $c_{1}=0$, from the dynamical system (\ref{ssf.37}) we
have the solution
\begin{equation}
x\left(  \tau\right)  =\left[  \left(  \frac{1}{\mu}+1\right)  x_{0}\tau
+x_{1}\right]  ^{\frac{\mu}{2\left(  1+\mu\right)  }}~,~y\left(  \tau\right)
=-\frac{\alpha}{\alpha^{2}-1}x\left(  \tau\right)  -y_{0}.
\end{equation}
where~$x_{0}=\frac{\varepsilon}{\alpha}\sqrt{2V_{0}\left(  \alpha
^{2}-1\right)  },$ and $\varepsilon=\pm1$.

In the case of $\mu=-1$ the solution of the system (\ref{ssf.37}) is%
\begin{equation}
x\left(  \tau\right)  =x_{1}e^{x_{0}\tau}~,~y\left(  \tau\right)
=-\frac{\alpha}{\alpha^{2}-1}x\left(  \tau\right)  -y_{0}%
\end{equation}
(recall that for $\mu=-1$, we have $dt=d\tau)$. Hence, the solution for the
scale factor is
\begin{equation}
a\left(  t\right)  =\left[  \frac{\alpha^{2}x_{1}^{2}}{\alpha^{2}-1}%
e^{2x_{0}t}-\frac{\alpha^{2}-1}{\alpha^{2}}y_{0}^{2}\right]  ^{\frac{1}{3}}\;.
\label{ssf.44}%
\end{equation}
%that is, we have an exponential solution.
Furthermore, from the singularity condition $a\left(  0\right)  =0$, we have
that $y_{0}^{2}=\frac{\alpha^{4}x_{1}^{2}}{\alpha^{2}-1}$ which implies
$a\left(  t\right)  =a_{1}\left(  e^{2x_{0}t}-1\right)  ^{\frac{1}{3}},~$where
$\alpha_{1}=\left(  \frac{\alpha^{2}x_{1}^{2}}{\alpha^{2}-1}\right)
^{\frac{1}{3}}$. We obtain $H\left(  t\right)  =\frac{2x_{0}}{3}%
\frac{e^{2x_{0}t}}{e^{2x_{0}t}-1},~$ and $t\left(  a\right)  =\frac{1}{2x_{0}%
}\ln\left[  1+\left(  \frac{a}{a_{1}}\right)  ^{3}\right]  $. Therefore, we
can express the Hubble function in terms of the scale factor, i.e.
\begin{equation}
E^{2}(a)=\frac{H^{2}(a)}{H_{0}^{2}}=\Omega_{\Lambda0}+{\tilde{\Omega}}%
_{m0}a^{-3}+\Omega_{sf0}a^{-6} \label{ssf.46}%
\end{equation}
where%
\[
\Omega_{\Lambda0}=\frac{4}{9}\frac{x_{0}^{2}}{H_{0}^{2}},~{\tilde{\Omega}%
}_{m0}=\frac{8}{9}\frac{x_{0}^{2}}{H_{0}^{2}}a_{1}^{3}~,~\Omega_{sf0}=\frac
{4}{9}\frac{x_{0}^{2}}{H_{0}^{2}}a_{1}^{6}%
\]
which means that the scalar field introduces an effective dark matter
component, namely ${\tilde{\Omega}}_{m}(a)={\tilde{\Omega}}_{m}a^{-3}%
/E^{2}(a)$ in the cosmic expansion.


\begin{thebibliography}{99}                                                                                               %


\bibitem {Teg04}M.~Tegmark \textit{et al.}, Astrophys.\ J.\ \textbf{606}, 702
(2004); D.~N.~Spergel \textit{et al.}, Astrophys.\ J.\ Supplt.\ \textbf{170},
377 (2007); T.~M.~Davis \textit{et al.}, Astrophys.\ J.\ \textbf{666}, 716
(2007); M.~Kowalski \textit{et al.}, Astrophys.\ J.\ \textbf{686}, 749(2008);
G.~Hinshaw \textit{et al.}, Astrophys.\ J.\ Supplt.\ \textbf{180}, 225 (2009);
J.~A.~S.~Lima and J.~S.~Alcaniz, Mon.\ Not.\ Roy.\ Astron.\ Soc.\ \textbf{317}%
, 893 (2000); J.~F.~Jesus and J.~V.~Cunha, Astrophys.\ J.\ Lett.\ \textbf{690}%
, L85 (2009); S.~Basilakos and M.~Plionis, Astrophys.\ J.\ Lett.\ \textbf{714}%
, 185 (2010); E. Komatsu E. et al., 2011, Astrophys.\ J.\ Sup., \textbf{192},
18 (2011); G. Hinshaw et al., Astrophys.\ J.\ Sup. \textbf{208}, 19 (2013); O.
Farooq, D. Mania and B. Ratra, Astrophys.\ J., \textbf{764}, 138 (2013); P. A.
R. Ade et al., (Planck Collaboration), Astronomy and Astrophysics
\textbf{571}, A16 (2014)

\bibitem {Ame10}
%L. Amendola and S. Tsujikawa, Dark Energy Theory and
%Observations (Cambridge University Press, Cambridge, England, 2010)
E. J. Copeland, M. Sami and S. Tsujikawa, Intern. Journal of Modern Physics D,
\textbf{15}, 1753,(2006); L. Amendola and S. Tsujikawa, \textit{{Dark Energy
Theory and Observations}}, Cambridge University Press, Cambridge UK, (2010).

\bibitem {Barrow}C. Rubano and J. D. Barrow, Phys. Rev. D. \textbf{64}, 127301 (2001)

\bibitem {Urena}L. A. Urena-Lopez, T. Matos, Phys. Rev. D \textbf{62,} 081302
(2000); V. Sahni and A. Starobinsky, Int. J. Mod. Phys. D \textbf{9} 373
(2000); J.A.E. Carrillo, J.M. Silva and J.A.S. Lima, Astr. Relativ. Astroph.:
New Phenomena and New States of Matter in the Universe, Proceedings of the
Third Workshop (IWARA07) (arXiv:0806.3299)

\bibitem {CapFru}S. Capozziello, N. Frusciante and D. Vernieri, Gen. Rel.
Grav. \textbf{44}, 1881 (2012)

\bibitem {Ritis}S. Capozziello and R. De Ritis, Phys. Lett. A., \textbf{203},
214; \textbf{203}, 283 (1995)

\bibitem {Capp96}S. Capozziello et al., Riv. Nuovo Cim., \textbf{19}, 1 (1996)

\bibitem {Cap96}S. Capozziello, E. Piedipalumbo, C. Rubano and P. Scudellaro,
Phys. Rev. D. \textbf{80} 104030 (2009); M. Szydlowski et al., Gen. Rel.
Grav., \textbf{38}, 795, (2006); \ Yi Zhang, Yun-gui Gong and Zong-Hong Zhu,
Phys. Lett. B., \textbf{688} 13 (2010); M. Sharif and I. Shafique, Phys. Rev.
D \textbf{90} 084033 (2014); D. Momeni, R. Myrzakulov and E. G\"{u}dekli
(arXiv:1502.00977); L. Collodel and G.M. Kremer, (arXiv:1411.3580)

\bibitem {Dong}H. Dong, J. Wang and X. Meng, Eur. Phys. J. C \textbf{73} 2543 (2013)

\bibitem {Darabi}F. Darabi, K. Atazadeh and A. Rezaei-Aghdam, Eur. Phys. J. C
\textbf{74} 2967 (2014)

\bibitem {Wei}H. Wei, X.J. Guo and L.F. Wang, Phys. Lett. B. \textbf{707} 298 (2012)

\bibitem {Kucu}Y. Kucukakca, U. Camci and I. Semiz, Gen.\ Relativ. Gravit.
\textbf{44} 1893 (2012); Y. Kucukakca, Eur. Phys. J. C \textbf{73} 2327
(2013); Y. Kucukakca, Eur. Phys. J. C \textbf{74} 3086 (2014)

\bibitem {CapHam}S. Capozziello, M. De Laurentis and S.D. Odintsov, Eur. Phys.
J. C \textbf{72} 2068 (2012)

\bibitem {Vak}Vakili B., Phys. Lett. B. \textbf{669} (2008) 206; B. Vakili, F.
Khazaie, Class. Quant. Grav. \textbf{29} 035015 (2012); B. Vakili, Phys. Lett.
B \textbf{738} 488 (2014)

\bibitem {Dim}N. Dimakis, T. Christodoulakis and P.A. Terzis, J. Geom. Phys.
\textbf{77} 97 (2014)

\bibitem {Dim2}P.A Terzis, N. Dimakis and T. Christodoulakis, Phys. Rev. D
\textbf{90} 123543 (2014)

\bibitem {Paliathanasis}A. Paliathanasis, M. Tsamparlis and S. Basilakos,
Phys. Rev. D. \textbf{84} 123514 (2011)

\bibitem {BasFT}A. Paliathanasis, M. Tsamparlis, S. Basilakos and S.
Capozziello, Phys. Rev. D. \textbf{89} 063532 (2014); A. Paliathanasis, S.
Basilakos, E.N. Saridakis, S. Capozziello, K. Atazadeh, F. Darabi and
M.\ Tsamparlis, Phys. Rev. D. \textbf{89} 104042 (2014); A. Paliathanasis and
M. Tsamparlis, Phys. Rev. D. \textbf{90} (2014) 043529; S. Basilakos, S.
Capozziello, M. De Laurentis, A. Paliathanasis and M. Tsamparlis, Phys. Rev.
D. \textbf{88} 103526 (2013)

\bibitem {DynSym}A. Paliathanasis, M. Tsamparlis and S. Basilakos, Phys. Rev.
D. \textbf{90} 103524 (2014); S. Basilakos, M. Tsamparlis and A.
Paliathanasis, Phys. Rev. D.\textbf{\ 83} 103512 (2011)

\bibitem {Kalotas}T.M. Kalotas and B.G. Wybourne, J. Phys. A: Math. Gen.
\textbf{15} 2077 (1982)

\bibitem {Maartens}R. Maartens and S.D. Maharaj, Class. WdWm Grav. \textbf{3}
1005 (1986)

\bibitem {TsamGRG}M. Tsamparlis and A. Paliathanasis, Gen. Relativ. Gravit.
\textbf{43} 1861 (2011)

\bibitem {AnIJGMMP}A. Paliathanasis and M. Tsamparlis, Int. J. Geom. Methods
Mod. Phys. \textbf{11} (2014) 1450037

\bibitem {Ellis05}J.R.\ Ellis, N.E. Mavromatos and D.V. Nanopoulos, Phys.
Lett. B \textbf{619} 17 (2005)

\bibitem {Ratra88}B. Ratra and P.J.E. Peebles, Phys. Rev. D \textbf{37} 3406 (1988)

\bibitem {Sievers}J.L. Sievers et al, Astrophys. J. \textbf{591} 599 (2003)

\bibitem {Bertacca}D. Bertacca, S. Matarrese and M. Pietroni, Mod. Phys. Lett.
A \textbf{22} 2893 (2007)

\bibitem {Brax}P. Brax and J. Martin, Phys. Lett. B \textbf{468} 40 (1999)

\bibitem {Gorini}V. Gorini, A. Kamenshchik, U. Moschella and V. Pasquier,
Phys. Rev. D \textbf{69} 123512 (2004)

\bibitem {Frieman}J.A. Frieman, C.T. Hill, A. Stebbins and I. Waga, Phys. Rev.
Lett. \textbf{75} 2077 (1995)

\bibitem {Sahni}V. Sahni and L.M. Wang, Phys. Rev. D \textbf{62} 103517 (2000)

\bibitem {BarrowLog}J.D. Barrow and P. Parsons, Phys.\ Rev. D \textbf{52} 5576 (1995)

\bibitem {Halliwel}J.J. Halliwell, Phys. Lett. B \textbf{185} 341 (1987)

\bibitem {Easther}R. Easther, Class. Quantum Grav. \textbf{10} 2203 (1993)

\bibitem {BarrowS}J.D. Barrow and P. Saich, Class. Quantum Grav. \textbf{10}
279 (1993)

\bibitem {Chimento}L.P. Chimento and A.E. Cossarini, Class. Quantum Grav.
\textbf{11} 1177 (1994)

\bibitem {Chimento2}L.P. Chimento and A. Jakubi, Int. J. Mod. Phys. D
\textbf{5} 71 (1996)

\bibitem {Ester2}E. Piedipalumbo, P. Scudellaro, G. Esposito and C. Rubano,
Gen. Relativ. Gravit. \textbf{44} 2611 (2012)

\bibitem {Bluman}G.W. Bluman and S. Kumei, \textit{Symmetries of Differential
Equations}, (Springer-Verlag, New York, (1989))

\bibitem {Ibrag}N.H. Ibragimov, Transformation groups applied to mathematical
physics, Translated from the Russian Mathematics and its Applications (Soviet
Series). D. Reidel Publishing Co., Dordrecht, (1985)

\bibitem {anthesis}A. Paliathanasis, Symmetries of differential equations and
applications in relativistic physics, PhD Thesis, University of Athens,
Athens, Greece (2014) (arXiv:1501.05129)

\bibitem {Arnold}V.I. Arnol'd, \textit{Mathematical Methods of Classical
Mechanics}, Graduate Texts in Mathematics, Vol. \textbf{60} (2nd ed.),
Springer (1989)

\bibitem {Saini00}T.~D.~Saini, S.~Raychaudhury, V.~Sahni, and
A.~A.~Starobinsky, Phys.\ Rev.\ Lett.\ \textbf{85}, 1162 (2000); D.~Huterer
and M.~S.~Turner, Phys.\ Rev.\ D \textbf{64} 123527 (2001).

\bibitem {Linjen03}E.~V.~Linder and A.~Jenkins,
Mon.\ Not.\ Roy.\ Astron.\ Soc.\ \textbf{346} 573 (2003); E.~V.~Linder,
Phys.\ Rev.\ D \textbf{72}, 043529 (2005)

\bibitem {copeland}E.J. Copeland, A.R. Liddle and D. Wands, Phys. Rev. D.
\textbf{57} 4686 (1998)

\bibitem {cop2}E.J. Copeland, S. Mizuno and M. Shaeri, Phys. Rev. D
\textbf{79} 103515 (2009)

\bibitem {pavl}S.A. Pavluchenko, Phys. Rev. D \textbf{67}, 103518 (2003)

\bibitem {Fadragas}C.R. Fadragas, G. Leon and E.N. Saridakis, Class. Quantum
Grav. \textbf{31} 075018 (2014)

\bibitem {Tamanini}N. Tamanini, Phys. Rev. D \textbf{89} 083521 (2014)

\bibitem {Ade15}P. A. R. Ade et al., (Planck Collaboration), (2015), [arXi:1502.01589]

\bibitem {HEW}R. N. Henriksen, A. Gordon Emslie, and P. S. Wesson Phys. Rev.
D., \textbf{27}, 1219 (1983)
\end{thebibliography}
\end{document}